\documentclass[twocolumn]{autart}    % Enable this line and disable the 
\usepackage{graphicx}          % Include this line if your 
\usepackage{xcolor}
\usepackage{url}
\usepackage{amsmath,amssymb,amsfonts}
\usepackage{subcaption}
\usepackage{algorithm}
\usepackage{algorithmic}

\usepackage{mathrsfs}
\usepackage{aligned-overset}
\usepackage{braket}
\usepackage{enumerate}
\usepackage[normalem]{ulem}

\newtheorem{theorem}{Theorem}
\newtheorem{lemma}{Lemma}
\newtheorem{definition}{Definition}
\newtheorem{remark}{Remark}
\newtheorem{example}{Example}
\begin{document}

\begin{frontmatter}

\title{Iterative Methods for Solving Feedback-Type Linear Equations under Small Phase Condition\thanksref{footnoteinfo}} % Title, preferably not more 

\thanks[footnoteinfo]{This work was supported in part by the National Natural Science Foundation of China under grants 62573006 and 72131001, by the Natural Science Foundation of Guangdong Province under grants 2024A1515011630, and by the Research Grants Council of Hong Kong under the General Research Fund No. 16206324. This paper was not presented at any IFAC 
meeting.}

\author[PKU]{Zhou Xie}\ead{zhouxie@pku.edu.cn},    % Add the 
\author[PKU]{Wei Chen\corauthref{cor1}}\ead{w.chen@pku.edu.cn},               % e-mail address 
\author[HKUSZ]{Li Qiu}\ead{qiuli@cuhk.edu.cn}  % (ead) as shown
\corauth[cor1]{Corresponding author.}
\address[PKU]{The School of Advanced Manufacturing and Robotics
\& the State Key Laboratory for Turbulence and Complex Systems, Peking
University, Beijing 100871, China}  % Please supply                                         
\address[HKUSZ]{The School of Science and Engineering, The Chinese
University of Hong Kong, Shenzhen, Shenzhen 518172, China}             % full addresses
        % here.

\begin{keyword}                           % Five to ten keywords,  
Feedback-type equation, numerical range, small phase theory, iterative method, 
numerical-radius contraction.             % chosen from the IFAC 
\end{keyword}                             % keyword list or with the 
                                          % help of the Automatica 
                                          % keyword wizard

\begin{abstract}                          % Abstract of not more than 200 words.
This paper investigates iterative methods for solving feedback-type linear equations under small phase condition. We first develop an iterative solution scheme for linear equations with sectorial (semi-sectorial) matrices, i.e., matrices whose numerical ranges are contained in an open (closed) half complex plane. Exploiting the geometric properties of the numerical range, we introduce the notion of numerical-radius contraction as a convergence criterion. Such a scheme, with some extra effort, leads to an iterative method for the feedback-type equations under small phase condition, whose convergence condition is characterized in terms of phase information. This complements classical iterative methods for feedback-type equations based on the small gain condition. A case study on large-scale power networks is presented to illustrate the computational efficiency of the proposed method.
\end{abstract}

\end{frontmatter}
\thispagestyle{plain}

\section{Introduction}
\begin{figure}[htbp]
	\centering
	\includegraphics[width=0.5\linewidth]{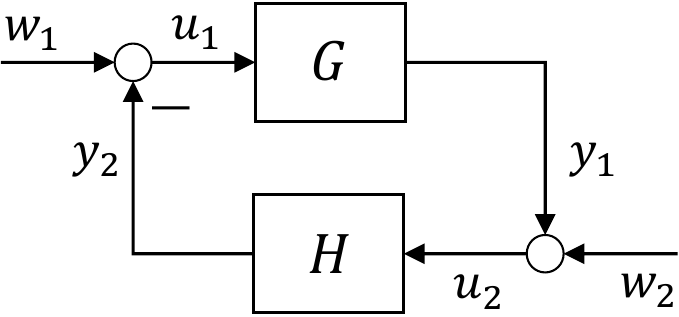}
	\caption{A feedback interconnection.}
	\label{fig: Standard feedback}
\end{figure}

Feedback mechanisms are fundamental to control theory and engineering, playing a pivotal role in ensuring stability, robustness, and performance guarantees \cite{aastrom2008feedback,doyle1992feedback,franklin2018feedback}. Many problems in feedback systems, such as evaluating closed-loop frequency response of a feedback interconnection as in Fig. \ref{fig: Standard feedback}, involve repeatedly solving linear equations of the form
\begin{equation}\label{eq: core_feedback_equation}
	\begin{bmatrix}
		I & H \\
		-G & I
	\end{bmatrix}\begin{bmatrix}
		u_1\\
		u_2
	\end{bmatrix}=\begin{bmatrix}
		w_1\\
		w_2
	\end{bmatrix},
\end{equation}
where $G,H\in\mathbb{C}^{n\times n}$ are determined by component frequency responses, and $u_1,u_2,w_1,w_2\in\mathbb{C}^n$.
This equation captures the input–output relation of the feedback interconnection.

The methods in solving linear equations can be broadly categorized into direct methods and iterative methods. The direct methods are designed to obtain solutions in finite steps, taking advantage of Gaussian elimination and its enhancements. They may face problems of robustness and expensive computational costs. Iterative methods, on the other hand, aim to approximate the solution through successive refinements, such that the refined iterations converge to the solution. Iterative methods are particularly favored in practical applications involving large-scale structured matrices due to their computational advantage \cite{kelley1995iterative,greenbaum1997iterative}.

A representative motivating example arises in the analysis of large-scale power networks. In this setting, $G$ is (block) diagonal, with each diagonal block capturing the local dynamics at an individual bus, while $H$ represents the coupling among buses induced by the underlying transmission network. Both $G$ and $H$ are typically large and sparse. This inherent sparsity makes iterative methods particularly well suited for solving the resulting equations.

The design and analysis of iterative methods for the feedback-type equation (\ref{eq: core_feedback_equation}) are closely tied to conditions that guarantee solvability. A classical result in this direction is the small gain theorem, which ensures invertibility of the feedback matrix $\begin{bmatrix} I & H \\
        -G & I\end{bmatrix}$ by requiring 
$\|G\|\|H\|<1$ \cite{zames1966input,stewart1990matrix}. This naturally leads to contractive fixed-point iterations \cite{zames1960nonlinear}. Nevertheless, as a norm-based condition, the small gain theorem can be overly conservative in situations where phase information plays a dominant role.
Recently, a small phase condition has been introduced as a counterpart to the small gain condition, which guarantees invertibility of the feedback matrix from a phasic point of view \cite{WCKQ2020,wang2023phases,Chen2024Phase}. The small phase theory was  built upon the concept of numerical range, a geometric object that encodes orientation information of a matrix. The small phase criterion is applicable to a broad range of engineering systems, including but not limited to positive real systems \cite{Kottenstette2014}, relaxation systems \cite{Willems1972}, and negative imaginary systems \cite{Petersen2010}. While the phase concept advances our understanding of the solvability of a feedback equation, a corresponding iterative solution method remains unexplored. 

This paper develops an iterative method to solve the feedback equation under small phase condition, thereby complementing the existing methods under small gain condition. The main contributions are threefold. First, we introduce the concept of numerical-radius contraction (numerical-radius semi-contraction, resp.) as a new convergence certificate motivated by the geometric properties of the numerical range. Numerical-radius contraction requires the numerical radius to be less than $1$, or equivalently, the numerical range lies within the open unit disk. Positioned between spectral-radius-based contraction and norm-based contraction, it gives a less conservative convergence condition than the latter while maintaining numerical robustness \cite{eiermann1993fields}. Second, based on numerical-radius contraction, we establish iterative schemes for solving linear equations with sectorial (semi-sectorial, resp.) matrices, i.e., matrices whose numerical ranges are contained in an open (closed, resp.) half-plane. Third, such iterative schemes are exploited to develop a method to solve feedback-type equation (1) under the small phase condition. A key ingredient is a novel convex-combination preconditioner that transforms the feedback-type equation (1) into one with a sectorial coefficient matrix. Compared with general-purpose iterative methods such as Krylov subspace methods, the proposed method exploits phase information to establish convergence guarantees and error bounds. Such convergence guarantees may remain applicable under matrix uncertainties preserving small phase condition.

The remainder of the paper is organized as follows. Section \ref{section: preliminaries} reviews some preliminaries on the numerical range and matrix phase. Section~\ref{section: NR_Contraction} discusses numerical-radius contraction and its role in the convergence of iterative methods. An iterative scheme is also given for linear equations with (semi-)sectorial matrices. Section \ref{section: linear equation case} presents the proposed iterative methods for feedback-type equations under the small phase condition. Section~\ref{section: example} provides a case study on power networks. The paper ends in Section \ref{section: conclusion} with concluding remarks.

\textit{Notation:} For a matrix $A\in\mathbb{C}^{n\times n}$, denote by $\mathcal{N}(A)$ the nullspace of $A$ and $\mathcal{R}(A)$ the range space. Denote by $A^*$ the conjugate transpose of $A$, $A^\dagger$ the Moore--Penrose inverse, and $\mathrm{Re}(A)=\frac{A+A^*}{2}$ the real part. An eigenvalue of $A\in\mathbb{C}^{n\times n}$ is said to be semi-simple if its algebraic and geometric multiplicity are equal. Let $\rho(A)$, $\Vert A\Vert$, $\Vert A\Vert_F$ denote the spectral radius, spectral norm, and Frobenius norm of $A\in\mathbb{C}^{n\times n}$, respectively. For a Hermitian matrix $A\in\mathbb{C}^{n\times n}$, denote its eigenvalues in non-increasing order by $\lambda_1(A)\geq \lambda_2(A)\geq \cdots\geq \lambda_n(A)$. We also use $\lambda_{\min}(A)$ ($\lambda_{\max}(A)$, resp.) to denote the smallest (largest, resp.) eigenvalue of a Hermitian $A$. We write $A\geq0$ ($A>0$, resp.)  to represent that $A$ is positive semidefinite (positive definite, resp.). Denote by $\mu(A)=\lambda_{\max}(\frac{A+A^*}{2})$ the log norm of a matrix $A\in \mathbb{C}^{n\times n}$. For a vector $x\in\mathbb{C}^n$, denote by $\Vert x\Vert=\sqrt{x^*x}$ the Euclidean norm of $x$. 
%For a nonzero complex number $z$ in the polar form $z=re^{j\phi}$ with $r>0$, denote by $\angle z=\phi$ the phase (also known as the argument) of $z$. Denote by $\overline{z}$ the conjugate of a complex number $z$.

\section{Preliminaries}\label{section: preliminaries}
We review some basics on numerical range and phases of a complex matrix. For more details, one can refer to \cite{horn1994topics} for numerical range and \cite{furtado2003spectral,WCKQ2020,wang2023phases} for matrix phases.

The numerical range (also known as the field of values) of a matrix $A\in \mathbb{C}^{n\times n}$ is defined as
\begin{equation*}
    W(A)=\left\{ x^* Ax:x\in\mathbb{C}^n,\Vert x\Vert=1 \right\}.
\end{equation*}
$W(A)$ is a compact and convex subset of $\mathbb{C}$, and contains the spectrum of $A$. For any $\alpha\in \mathbb{C}$ and $A,B\in \mathbb{C}^{n\times n}$, the following properties of the numerical range are immediate: (i) homogeneity: $W(\alpha A)=\alpha W(A)$; (ii) translation property: $W(A+\alpha I)=W(A)+\left\{\alpha\right\}$; (iii) subadditivity: $W(A+B)\subset W(A)+W(B)$. 
% \begin{enumerate}[(i)]
%     \item Homogeneity: $W(\alpha A)=\alpha W(A)$;
%     \item Translation property: $W(A+\alpha I)=W(A)+\left\{\alpha\right\}$;
%     \item Subadditivity: $W(A+B)\subset W(A)+W(B)$.                  
%     % \item Unitary invariance: $W(U^*AU)=W(A)$ for any unitary matrix $U\in\mathbb{C}^{n\times n}$.
% \end{enumerate}
It is not hard to show that
\begin{align*}
    \min_{z\in W(A)} \mathrm{Re}(z) &=\lambda_{\min}(\mathrm{Re}(A)),\\
    \max_{z\in W(A)} \mathrm{Re}(z) &=\lambda_{\max}(\mathrm{Re}(A))=\mu(A),
\end{align*}
meaning $W(A)$ lies between two vertical lines that intersect the real axis at $\lambda_{\min}(\mathrm{Re}(A))$ and $\mu(A)$ respectively.

The numerical radius of a matrix $A\in\mathbb{C}^{n\times n}$ is defined as $ w(A)=\max\{|z|:z\in W(A)\}.$
% \begin{equation*}
%     w(A)=\max\{|z|:z\in W(A)\}.
% \end{equation*}
The following lemmas are useful in later developments.
\begin{lemma}[\cite{horn1994topics}]\label{lemma:numericalradius}
$\rho(A)\leq w(A)\leq \Vert A\Vert \leq 2w(A).$
    % \[
    % \rho(A)\leq w(A)\leq \Vert A\Vert \leq 2w(A).
    % \]
\end{lemma}

\begin{lemma}[\cite{henrici1962bounds}]\label{lemma:depature_from_normality}
$w(A)-\rho(A)\leq\sqrt{\frac{1-n^{-1}}{2}}\,\Delta(A)$,
where $\Delta(A):=\sqrt{\|A\|_F^2-\sum_{i=1}^n|\lambda_i(A)|^2}$
measures the departure from normality.
\end{lemma}

\begin{lemma}[\cite{berger1965strange}]\label{lemma: power_ineq}
$w(A^k)\leq w(A)^k$ for any integer $k\geq 0$.
    % For any integer $k\geq 0$, there holds 
    % $
    %     w(A^k)\leq w(A)^k.
    % $
\end{lemma}

The angular numerical range of $A\in\mathbb{C}^{n\times n}$, denoted by $W_a(A)$, is defined as the conic hull of $W(A)$, i.e.,
\begin{equation*}
    W_a(A)=\left\{ x^* Ax:x\in\mathbb{C}^n, x\neq 0\right\}.
\end{equation*}
The field angle of $A$, denoted by $\delta(A)$, is defined as the angle subtended by $W_a(A)$ if $W_a(A)$ does not contain a line through the origin, as $\pi$ if $W_a(A)$ contains one line through the origin, as $2\pi$ if $W_a(A)$ is the whole complex plane. 
%We refer readers to \cite{horn1991topics} for more details on numerical range.   
\begin{definition}[\cite{wang2023phases}]
A matrix $A\!\in\!\mathbb{C}^{n\times n}$ is said to be 
\begin{enumerate}[(i)]
    \item sectorial if $0\notin W(A)$;
    \item quasi-sectorial if $\delta(A)<\pi$;
    \item semi-sectorial if $\delta(A)\leq \pi$.
\end{enumerate}
\end{definition}
Clearly, for a sectorial matrix $A$, $W(A)$ is contained in an open half plane due to its convexity, and hence $\delta(A)<\pi$. Thus, $ \{\text{sectorial matrices} \} \subset \{\text{quasi-sectorial matrices} \}\subset \{\text{semi-sectorial matrices}\},$
% \begin{align*}
% \{\text{sectorial matrices} \}&\subset \{\text{quasi-sectorial matrices} \} \\
% &\qquad\qquad\subset \{\text{semi-sectorial matrices} \},
% \end{align*}
where the set containments are proper ones. 

\begin{figure}[htp]
\centering
\subfloat[Sectorial matrix]{%
  \begin{minipage}{0.45\linewidth}
    \centering
    \includegraphics[width=0.75\linewidth]{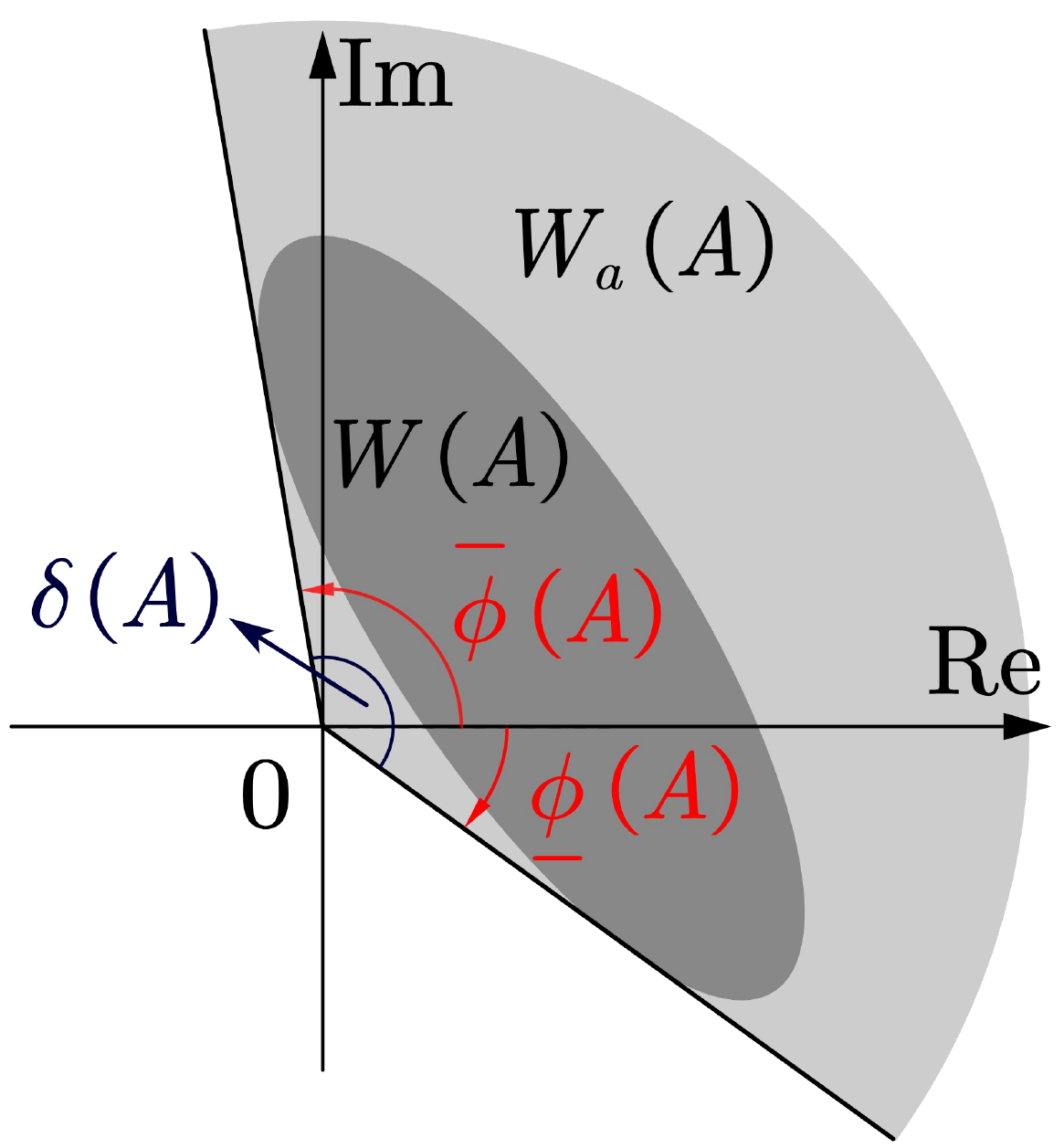}
  \end{minipage}
  \label{fig: sectorial}
}
\hfil
\subfloat[Quasi-sectorial matrix]{%
  \begin{minipage}{0.45\linewidth}
    \centering
    \includegraphics[width=0.75\linewidth]{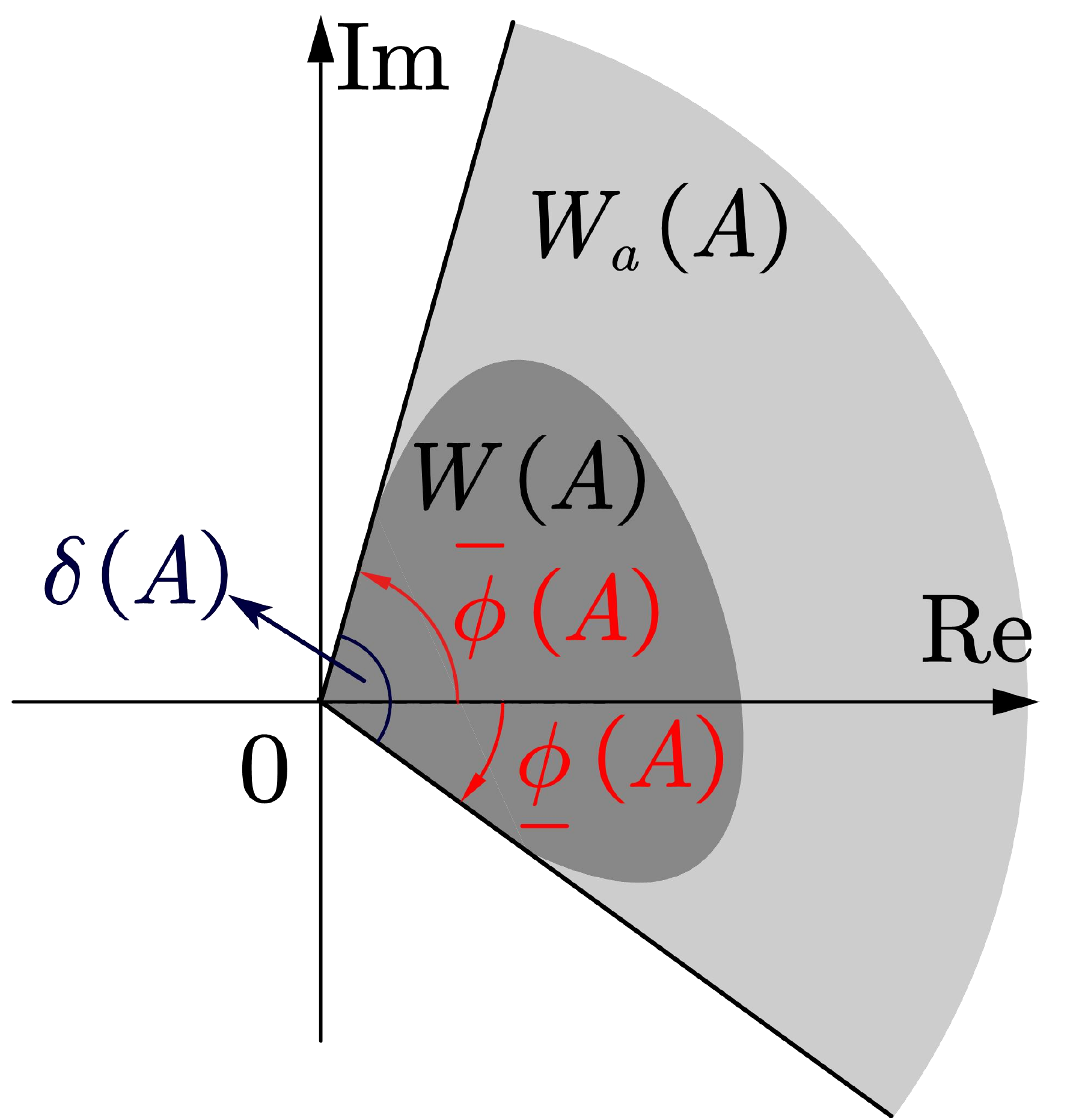}
  \end{minipage}
  \label{fig: quasi-sectorial}
}
\vspace{10pt}
\subfloat[Semi-sectorial matrix]{%
  \begin{minipage}{0.45\linewidth}
    \centering
    \includegraphics[width=0.75\linewidth]{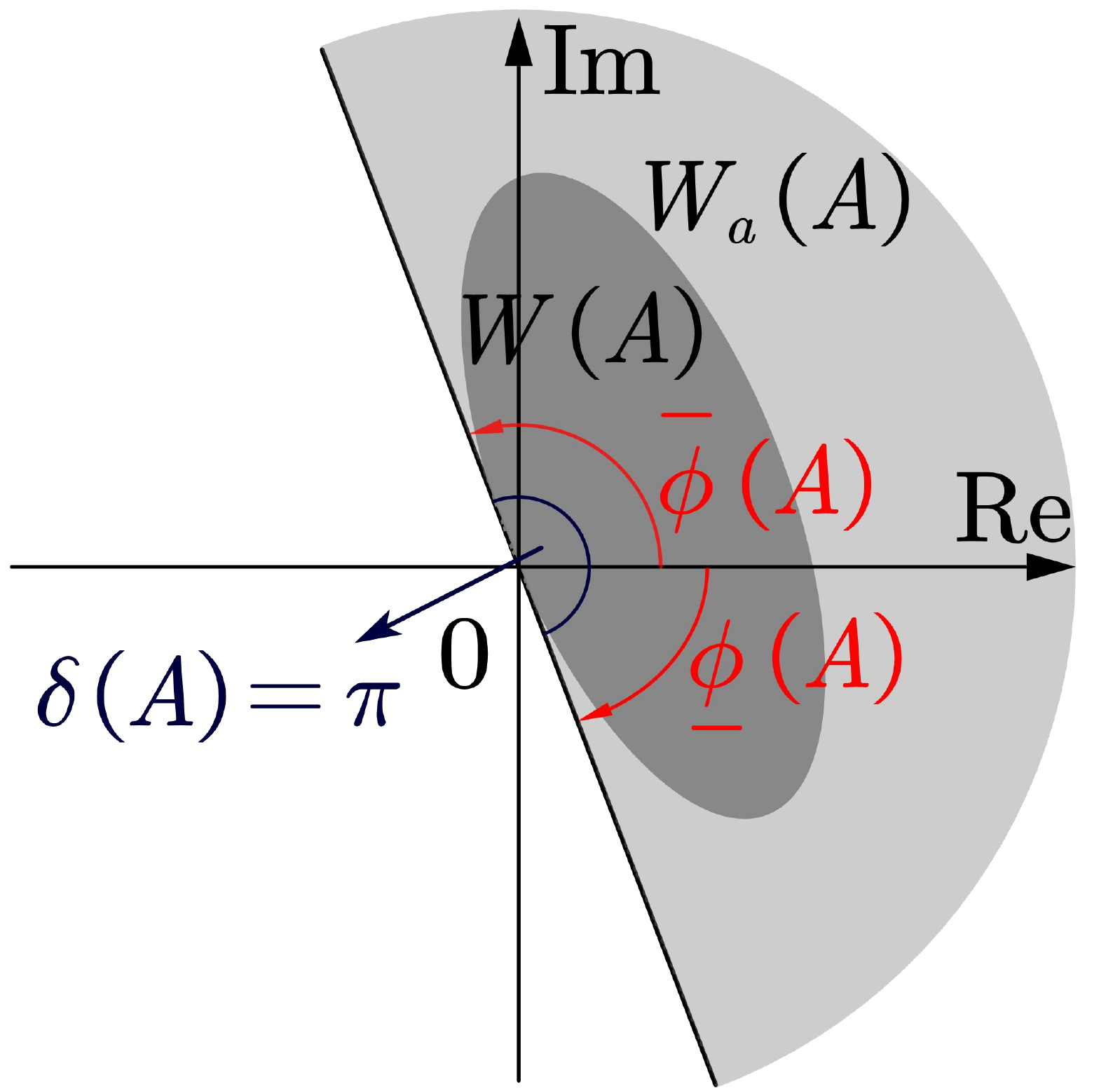}
  \end{minipage}
  \label{fig: semi-sectorial}
}
\hfil
\subfloat[Non-semi-sectorial matrix]{%
  \begin{minipage}{0.48\linewidth}
    \centering
    \includegraphics[width=0.75\linewidth]{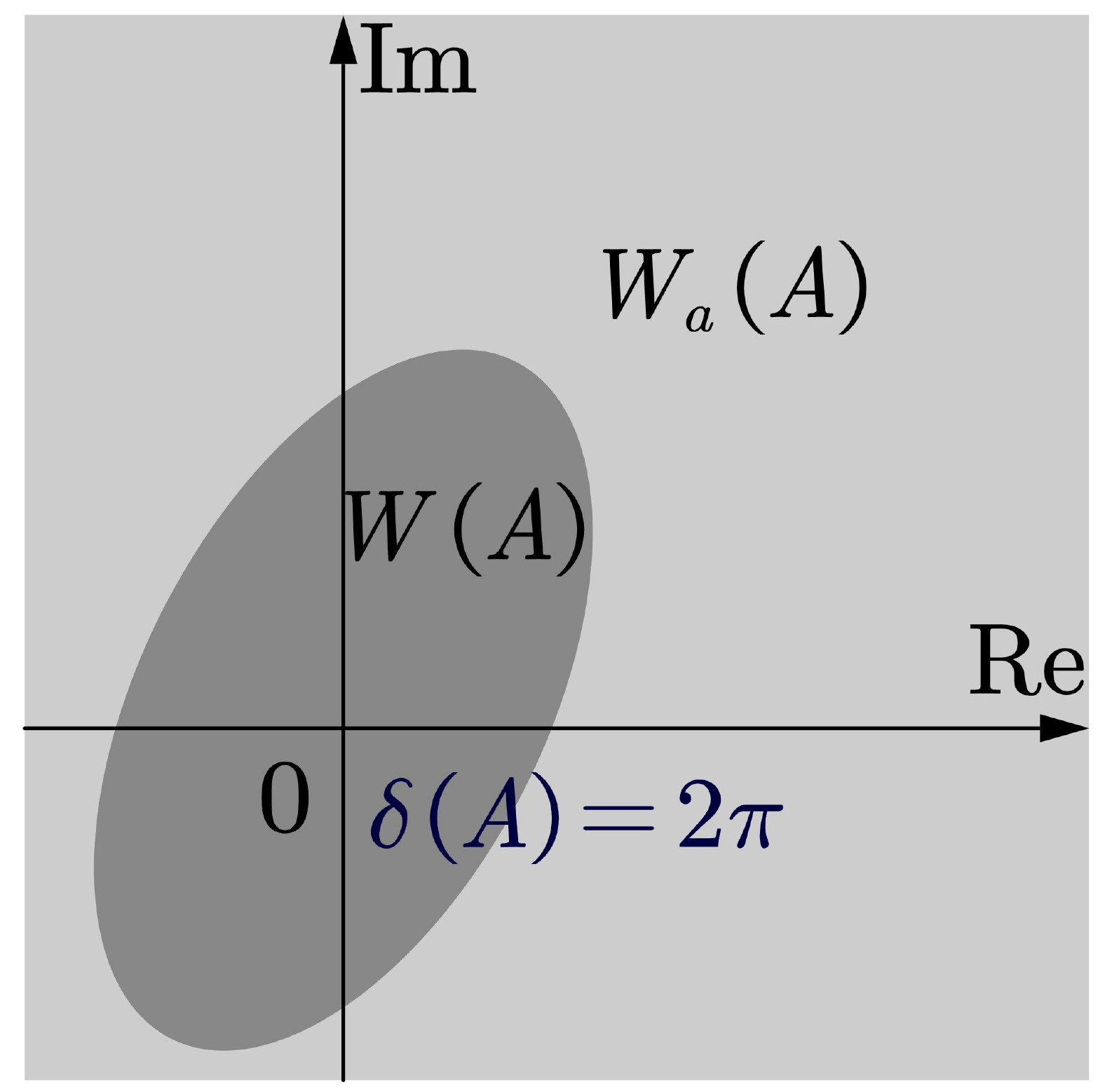}
  \end{minipage}
  \label{fig: non-semi-sectorial}
}
\caption{Illustrations of sectorial, quasi-sectorial, semi-sectorial, and non-semi-sectorial matrices, where $W(A)$ and $W_a(A)$ are represented by dark gray and light gray regions. The phases of respective matrices are also illustrated.}
\label{fig: illustration of sectorial}
\end{figure}

For a semi-sectorial matrix $A\in\mathbb{C}^{n\times n}$, the minimum phase and the maximum phase of $A$, denoted by $\underline{\phi}(A)$ and $\overline{\phi}(A)$, respectively, are defined as \cite{wang2023phases}
\begin{align*}
        \underline{\phi}(A)&=\inf_{z\in W(A)\backslash\{0\}} \angle z, \\
        \overline{\phi}(A)&=\sup_{z\in W(A)\backslash\{0\}}  \angle z.
\end{align*} 
The values of minimum and maximum phases are in general determined modulo $2\pi$. Once we choose a value of $\gamma(A)=\frac{\overline{\phi}(A)+\underline{\phi}(A)}{2},$
% \[
% \gamma(A)=\frac{\overline{\phi}(A)+\underline{\phi}(A)}{2},
% \]
called the phase center of $A$, in $\mathbb{R}$, then $\underline{\phi}(A)$ and $\overline{\phi}(A)$ are uniquely determined. In addition, the minimum and maximum phases are said to take principal values if $\gamma(A)$ is chosen in $[-\pi,\pi)$. It is clear that the field angle $\delta(A)=\overline{\phi}(A)-\underline{\phi}(A).$ When $W(A)=\{0\}$, $A$ is a zero matrix and is understood to be quasi-sectorial. In this case, $W(A)\backslash\{0\}=\emptyset$ and following convention  we let  $\underline{\phi}(0)=\infty,\overline{\phi}(0)=-\infty$. An illustration of numerical ranges and phases of matrices is shown in Fig. \ref{fig: illustration of sectorial}. For a semi-sectorial $A$, $\underline{\phi}(A)$ and $\overline{\phi}(A)$ can be computed in polynomial time using LMI solvers \cite{wang2023phases}. For simplicity, we assume that the phases take principal values throughout the paper.

For a semi-sectorial matrix $A\in\mathbb{C}^{n\times n}$ with rank $r$, it has a decomposition \cite{furtado2003spectral}
\begin{equation}\label{eq: sectorial_decomposition}
    A=U\begin{bmatrix}
        0 & 0 \\
        0 & A_s
    \end{bmatrix}U^*,
\end{equation}
where $U\in \mathbb{C}^{n\times n}$ is unitary and $A_s\in\mathbb{C}^{r\times r}$ is nonsingular semi-sectorial. In addition, $A$ is quasi-sectorial if and only if $A_s$ is sectorial \cite{chien1998geometric,wang2023phases}. From decomposition (\ref{eq: sectorial_decomposition}), we can see that $\mathcal{N}(A)=\mathcal{N}(A^*)$ and $\mathcal{N}(A)\perp \mathcal{R}(A)$. Moreover, $W(A)$ is the convex hull of $\{0\} \cup W(A_s)$. Thus 
%$\underline{\phi}(A)=\underline{\phi}(A_s)$ and $\overline{\phi}(A)=\overline{\phi}(A_s)$.
\begin{align}
\underline{\phi}(A)=\underline{\phi}(A_s) \text{ and } \overline{\phi}(A)=\overline{\phi}(A_s).\label{phasecompression}
\end{align}

A matrix $A\!\in\!\mathbb{C}^{n\times n}$ is said to be accretive (strictly accretive, resp.) if $\mathrm{Re}(A)\geq 0$ ($\mathrm{Re}(A)> 0$, resp.) \cite{kato1980perturbation}. By definition, a matrix is accretive (strictly accretive, resp.) if and only if it is semi-sectorial (sectorial, resp.) with phases taking principal values in $[-\frac{\pi}{2},\frac{\pi}{2}]$ ($(-\frac{\pi}{2},\frac{\pi}{2})$, resp.). We define a quasi-strictly accretive matrix as a quasi-sectorial matrix with phases taking principal values in $(-\frac{\pi}{2},\frac{\pi}{2})$. 
The following lemma is useful.

\begin{lemma}\label{lemma: when_semi_to_quasi}
    Let $A\in\mathbb{C}^{n\times n}$ be accretive. Then $A$ is further quasi-strictly accretive if and only if $\mathcal{N}(A)=\mathcal{N}(\mathrm{Re}(A))$.
\end{lemma}
\begin{pf}
The ``only if'' part was shown in \cite[p.~135]{arlinskii2003sectorial}. An alternative proof is provided here for the sake of completeness. For an accretive $A$, it admits the decomposition (\ref{eq: sectorial_decomposition}) and thus $\mathcal{N}(A)\subset \mathcal{N}(\mathrm{Re}(A))$. If $A$ is quasi-strictly accretive, then $A_s$ as in the decomposition (\ref{eq: sectorial_decomposition}) is strictly accretive and therefore $\mathrm{rank}(A)=\mathrm{rank}(\mathrm{Re}(A))$. This shows $\mathcal{N}(A)=\mathcal{N}(\mathrm{Re}(A))$.

Now we show ``if'' part. Since $A$ is accretive, it admits the decomposition (\ref{eq: sectorial_decomposition}) with $A_s$ nonsingular accretive. It follows that $\mathrm{Re}(A)=U\begin{bmatrix}
        0 & 0 \\
        0 & \mathrm{Re}(A_s)
    \end{bmatrix}U^*$ with $\mathrm{Re}(A_s)\geq 0$. If further $\mathcal{N}(A)=\mathcal{N}(\mathrm{Re}(A))$, then $\mathrm{Re}(A_s)$ is nonsingular, and hence $\mathrm{Re}(A_s)>0$, indicating that $A_s$ is strictly accretive and $A$ is quasi-sectorial. In light of (\ref{phasecompression}), $\underline{\phi}(A)$ and $\overline{\phi}(A)$ take principal values in $(-\frac{\pi}{2},\frac{\pi}{2})$. The proof is complete.
\end{pf}

The following small phase theorem is shown in \cite{wang2023phases}.
\begin{lemma}\label{lemma:small_phase_condition}
    Let $G,H\in\mathbb{C}^{n\times n}$ be semi-sectorial matrices. Then $\begin{bmatrix}
        I & H \\
        -G & I
    \end{bmatrix}$ has full rank if 
    \begin{align}\label{eq: matrix_small_phase_condition}
          \overline{\phi}(G)+\overline{\phi}(H)<\pi \text{ and }
                  \underline{\phi}(G)+\underline{\phi}(H)>-\pi.
    \end{align}
\end{lemma}

The small phase condition (\ref{eq: matrix_small_phase_condition}) provides a phasic counterpart to the well known small gain condition as follows \cite{stewart1990matrix}. They establish nonsingularity of feedback matrix $\begin{bmatrix}
        I & H \\
        -G & I
    \end{bmatrix}$ using phase information and gain information respectively.

% The small phase theorem is a parallel result to the celebrated matrix small gain theorem as follows \cite{stewart1990matrix}.
\begin{lemma}\label{lemma: small gain condition}
    Let $G,H\in\mathbb{C}^{n\times n}$. Then $\begin{bmatrix}
         I & H \\
         -G & I
     \end{bmatrix}$ has full rank if $\Vert G\Vert \cdot \Vert H\Vert<1$.
\end{lemma}

\section{Numerical-radius Contraction and Convergence of Iterative Methods}\label{section: NR_Contraction}
\subsection{Numerical-radius contraction and semi-contraction}

\begin{definition}[\cite{FB-LNS}]
A matrix $A\!\in\!\mathbb{C}^{n\times n}$ is said to be semi-convergent if $\lim_{k\rightarrow +\infty}A^k$ exists, and convergent if further $\lim_{k\rightarrow +\infty}A^k=0$.
\end{definition}

For an $A\in\mathbb{C}^{n\times n}$, it is convergent if and only if $\rho(A)<1$;it is semi-convergent but not convergent if and only if $1$ is a semi-simple eigenvalue of $A$ and all other eigenvalues have magnitude less than $1$ \cite{FB-LNS}.

It is known that $\rho(A)$ is highly sensitive to perturbations. On the contrary, the norm $\|A\|$ is numerically robust and forms the foundation of classical contraction theory \cite{FB-CTDS}, which has been widely employed in convergence analysis of numerical algorithms, particularly in fixed-point iterations. As shown in Lemma \ref{lemma:numericalradius}, the numerical radius $w(A)$ provides a tighter upper bound on the spectral radius than the norm. In addition, it is numerically reliable since $W(A+\Delta)\subset W(A)+W(\Delta)\subset W(A)+\left\{z\in \mathbb{C}:|z|\!\leq\! \Vert \Delta\Vert \right\}$
% \begin{multline*} 
% W(A+\Delta)\subset W(A)+W(\Delta)\\
% \subset W(A)+\left\{z\in \mathbb{C}:|z|\!\leq\! \Vert \Delta\Vert \right\} 
% \end{multline*} 
for any perturbation $\Delta\!\in\!\mathbb{C}^{n\times n}$. This suggests that the numerical radius offers a less conservative measure than the matrix norm in convergence analysis, while maintaining robustness to perturbations.

We introduce the following definition of numerical-radius contractivity and semi-contractivity.

\begin{definition}
    A matrix $A\in \mathbb{C}^{n\times n}$ is said to be numerical-radius contractive ($w$-contractive) if $w(A)<1$, and is said to be numerical-radius semi-contractive ($w$-semi-contractive) if $w(A)\leq 1$ and $W(A)$ intersects the unit circle at most at point $1$.
\end{definition}

\begin{example}
Consider matrix  $A=\begin{bmatrix}
0.2&1.2\\
0.3&0.2
\end{bmatrix},$ for which $w(A)=0.95<1$ and $\|A\|\approx1.24>1$. Thus, $A$ is numerical-radius contractive but not norm-contractive. 
\end{example}

\begin{figure}[htbp]
\centering
\subfloat[]{%
  \begin{minipage}{0.3\linewidth}
    \centering
    \includegraphics[width=1.0\linewidth]{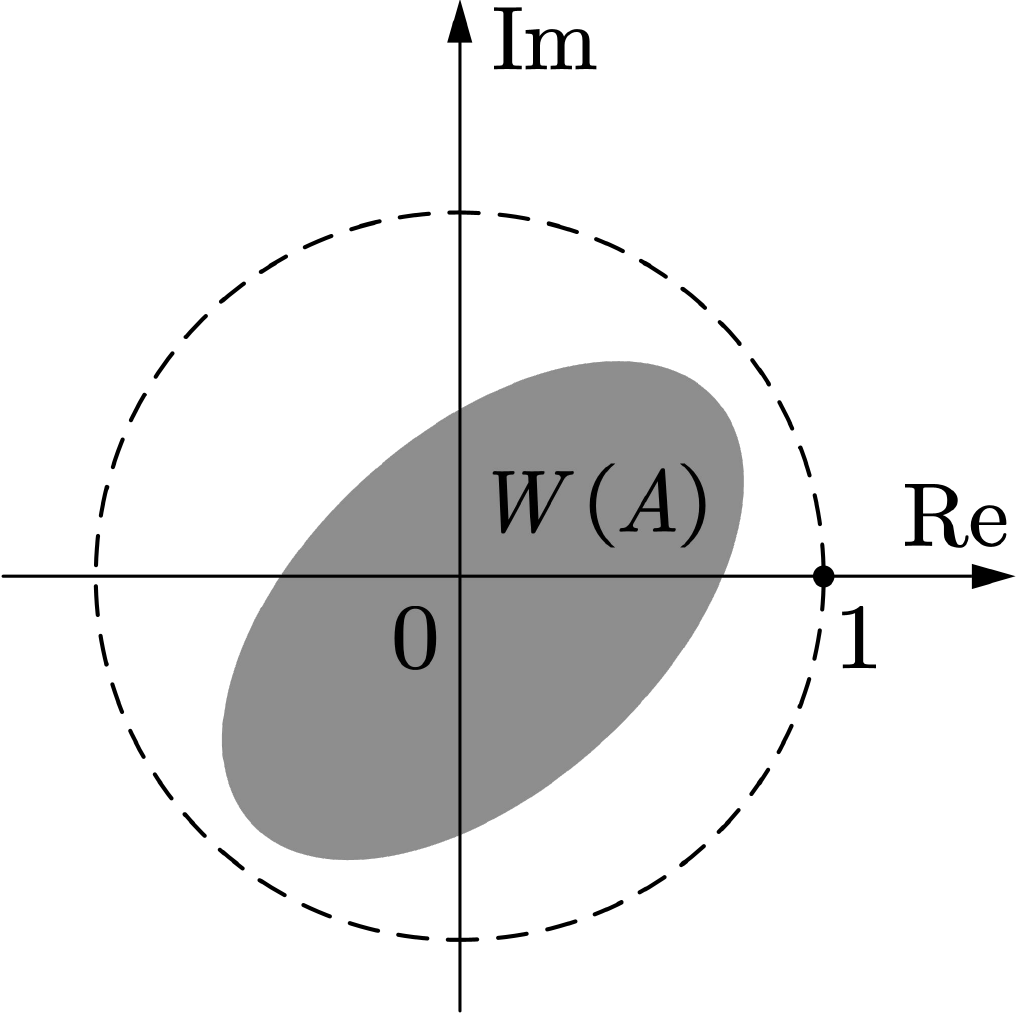}
  \end{minipage}
  \label{fig: NR_contractive}
}
\hfil
\subfloat[]{\begin{minipage}{0.3\linewidth}
    \centering
    \includegraphics[width=1.0\linewidth]{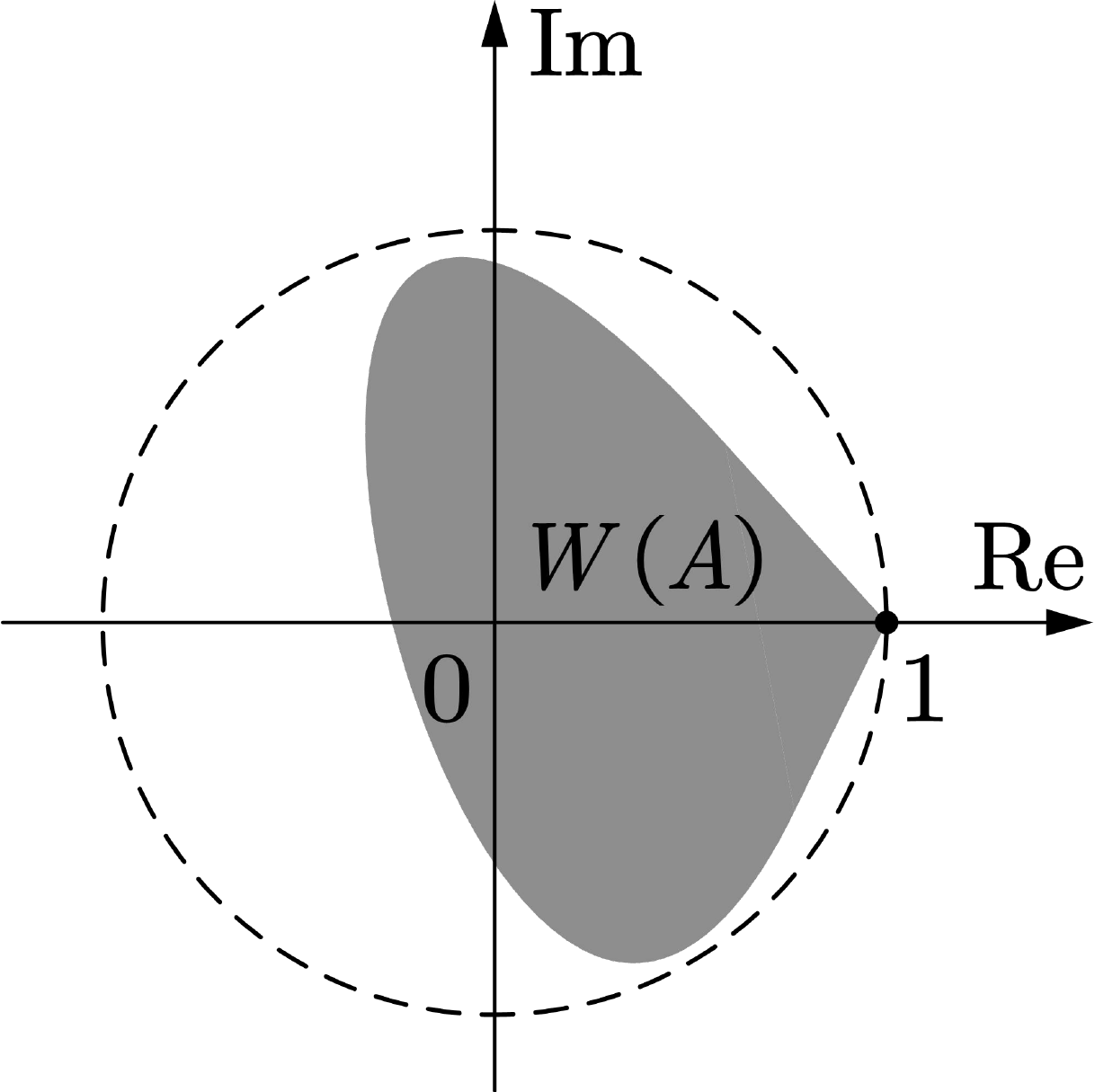}
  \end{minipage}
  \label{fig: NR_semicontractive_sharp}
}
\hfil
\subfloat[]{
  \begin{minipage}{0.3\linewidth}
    \centering
    \includegraphics[width=1.0\linewidth]{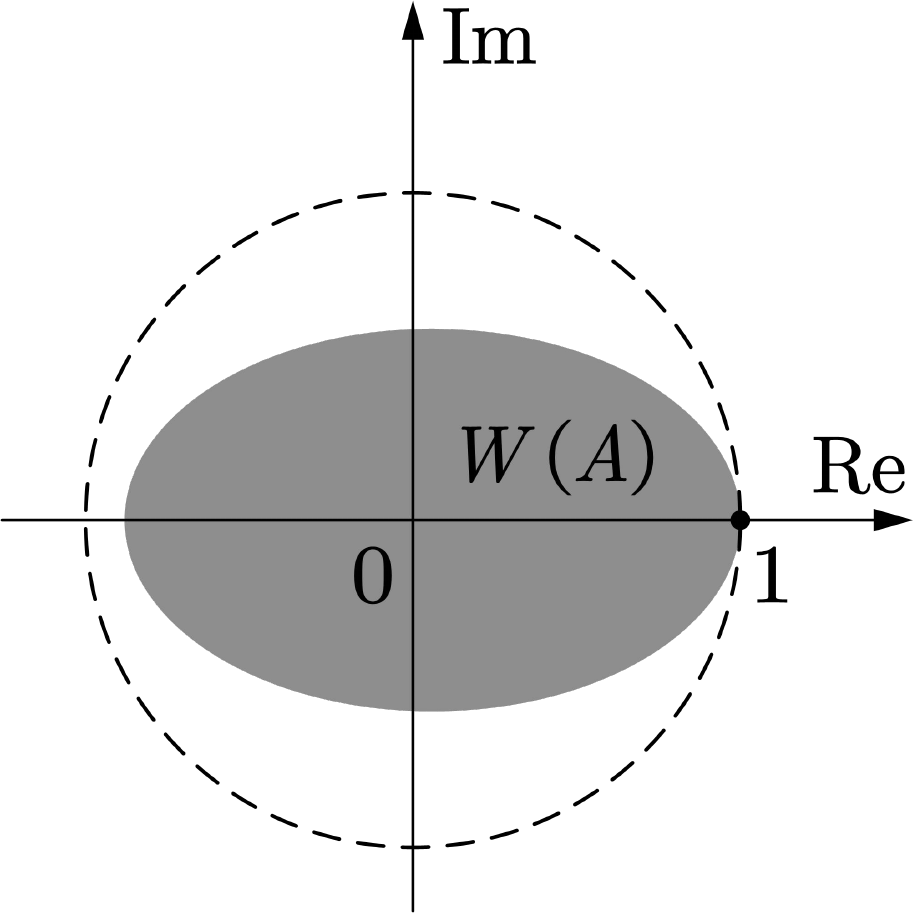}
  \end{minipage}
  \label{fig: NR_semicontractive_smooth}
}
%\caption{$w$-contractive and $w$-semi-contractive matrices.}
\caption{$w$-contractive and $w$-semi-contractive matrices:
(a) $w$-contractive;
(b) $w$-semi-contractive ($1$ is a sharp point of $W(A)$);
(c) $w$-semi-contractive ($1$ is on the smooth boundary of $W(A)$).}
\label{fig: NR_contractive matrix}
\end{figure}

Clearly, $w$-contractivity ($w$-semi-contractivity, resp.) of $A$ implies convergence (semi-convergence, resp.) of $A$. A graphical illustration of $w$-contractive and $w$-semi-contractive matrices is shown in Fig. \ref{fig: NR_contractive matrix}. 

From the graphical illustration, one can see that for $w$-semi-contractive matrices, two cases may occur: (i) $1$ is a sharp point of $W(A)$ as in Fig. \ref{fig: NR_contractive matrix}(b); (ii) $1$ lies on the smooth boundary of $W(A)$ as in Fig. \ref{fig: NR_contractive matrix}(c).
% \begin{enumerate}[1)]
%     \item $1$ is a sharp point of $W(A)$ as in Fig. \ref{fig: NR_contractive matrix}(b);
%     \item $1$ lies on the smooth boundary of $W(A)$ as in Fig. \ref{fig: NR_contractive matrix}(c).
% \end{enumerate}
In the first case, by \cite[Theorem 1.6.3, Theorem 1.6.6]{horn1994topics}, $1$ is a normal eigenvalue of $A$, namely, a semi-simple eigenvalue whose eigenspace is orthogonal to the eigenspaces of all other eigenvalues. Hence $A$ has a decomposition 
\begin{align}\label{wsemicontr}
    A=V\begin{bmatrix}I&0\\0&A_c\end{bmatrix}V^*,
\end{align}
where $V$ is a unitary matrix and $A_c$ is $w$-contractive. In the second case, $1$ may or may not be an eigenvalue of $A$. If $1$ is indeed an eigenvalue, then again by \cite[Theorem 1.6.6]{horn1994topics}, it is a normal eigenvalue, and $A$ admits a decomposition in the form of (\ref{wsemicontr}), where $A_c$ is $w$-semi-contractive with no eigenvalues at $1$.

The following lemma shows the role of $w$-contractivity and $w$-semi-contractivity in the convergencethe  of iteration
\begin{equation}\label{eq: A-iteration}
x(k+1)=Ax(k)+b,\quad x(0)\in \mathbb{C}^n,
\end{equation}
where $A\in \mathbb{C}^{n\times n}$ and $b\in\mathbb{C}^n$. Assume that $b\in\mathcal{R}(I-A)$. This ensures the existence of fixed points of (\ref{eq: A-iteration}).
\begin{lemma}\label{lemma: NR_contractive_convergence}
    The following statements hold.
\begin{enumerate}[(i)]
    \item If $A$ is $w$-contractive, then iteration \eqref{eq: A-iteration} converges to $x^\star=(I-A)^{-1}b$ with an error bound 
    \begin{equation}\label{eq: w-contractive_error_bound}
    \Vert x(k)-x^\star\Vert \leq 2w(A)^k \Vert x(0)-x^\star\Vert.
    \end{equation}
    \item  If $A$ is $w$-semi-contractive, then iteration \eqref{eq: A-iteration} converges to $x^\star=V_1V_1^*x(0)+(I-A)^{\dagger}b$, where $V_1$ is an isometry whose columns form a basis of $\mathcal{N}(I-A)$. Furthermore, if $1$ is a sharp point of $W(A)$, then 
    \begin{equation}\label{eq: NR_error_bound}
    \Vert x(k) -x^\star\Vert\leq 2w\left(A_c\right)^k \Vert x(0)-x^\star \Vert,
    \end{equation}
    where $A_c={V_1^\perp}^* AV_1^\perp$ with $V_1^\perp$ being an orthogonal complement of $V_1$.
\end{enumerate}
\end{lemma}
\begin{pf}
(i): 
Since $A$ is $w$-contractive, $A$ is convergent and iteration \eqref{eq: A-iteration} converges to $x^\star=\lim_{k\rightarrow +\infty}x(k)=(I-A)^{-1}b$. 
In addition, the error dynamics is given by $x(k+1)-x^\star = A(x(k)-x^\star).$
The error bound then follows from
\begin{align*}
    \Vert x(k)-x^\star\Vert&=\Vert A^k (x(0)-x^\star)\Vert \leq \Vert A^k \Vert\cdot \Vert x(0)-x^\star\Vert \\
    &\leq 2w(A^k) \Vert x(0)-x^\star\Vert \leq 2w(A)^k \Vert x(0)-x^\star\Vert,
\end{align*}
where the second inequality is due to Lemma \ref{lemma:numericalradius} and the last inequality is due to Lemma \ref{lemma: power_ineq}.

(ii): Since $A$ is $w$-semi-contractive, it can be decomposed as in (\ref{wsemicontr}), where $V=\begin{bmatrix}V_1&V_1^\perp\end{bmatrix}$ and $A_c={V_1^\perp}^* AV_1^\perp$. Moreover, $A_c$ is $w$-semi-contractive with no eigenvalues equal to $1$. This implies $\rho(A_c)<1$, i.e., $A_c$ is convergent. Now let $x_{//}(k)=V_1^*x(k)$ and $x_{\perp}(k)={V_1^\perp}^*x(k)$. Then iteration~\eqref{eq: A-iteration} decomposes as
\[
\begin{cases}
    x_{//}(k+1)=x_{//}(k),\\
    x_{\perp}(k+1)=A_c x_{\perp}(k)+{V_1^{\perp}}^*b.
\end{cases}
\]
Since $A_c$ is convergent, we have
\[
\begin{cases}
    x_{//}(k)=x_{//}(0),\quad \forall k\in \mathbb{Z}_{\geq 0}, \\
    \lim\limits_{k\rightarrow +\infty}x_{\perp}(k)=(I-A_c)^{-1}{V_1^{\perp}}^*b,
\end{cases}
\]
yielding that 
\begin{align}
    x^\star&=\lim\limits_{k\rightarrow\infty} x(k)=\lim\limits_{k\rightarrow\infty}VV^*x(k) \nonumber\\
    &=V_1V_1^*x(0)+V_1^\perp (I-A_c)^{-1}{V_1^{\perp}}^*b.\label{xstar}
\end{align}
Further, in view of decomposition~\eqref{wsemicontr}, we have
\[
I-A=V\begin{bmatrix}
    0 & 0 \\
    0 & I-A_c
\end{bmatrix}V^*=V_1^\perp (I-A_c){V_1^{\perp}}^*.
\]
This means $I-A$ admits a full rank decomposition $I-A=FG$, where $F=V_1^\perp (I-A_c)$ and $G={V_1^{\perp}}^*$. By \cite[Theorem 1.5]{BenIsraelGreville2003},
\[
(I\!-\!A)^\dagger=G^*\left( F^*(I\!-\!A)G^*  \right)^{-1}F^*=V_1^\perp (I\!-\!A_c)^{-1}{V_1^{\perp}}^*.
\]
In view of (\ref{xstar}), we have $x^\star=V_1V_1^*x(0)+(I-A)^{\dagger}b$.

If, in addition, $1$ is a sharp point of $W(A)$, then $A_c$ is $w$-contractive. Define $x_\perp^\star=\lim\limits_{k\rightarrow +\infty}x_{\perp}(k)$. Note that 
\begin{align*}
    &\Vert x(k)-x^\star\Vert=\left\Vert \begin{bmatrix}
        V_1 & V_1^\perp
    \end{bmatrix}^* (x(k)-x^\star) \right\Vert \\
    &=\left\Vert  \begin{bmatrix}
        x_{//}(k) \\
        x_\perp(k)
\end{bmatrix}-\begin{bmatrix}
        x_{//}(k) \\
        x_\perp^\star
    \end{bmatrix} \right\Vert=\Vert x_\perp(k)-x_\perp^\star \Vert.
\end{align*}

 The error bound then follows from an argument similar to that in part~(i).
\end{pf}

\begin{remark}
If $1$ lies on the smooth boundary of $W(A)$, $w(A_c)$ may equal $1$, rendering the error bound in \eqref{eq: NR_error_bound} trivial. Nevertheless, linear convergence can still be established under a suitable weighted norm. Since $\rho(A_c) < 1$, following a similar argument as in \cite[Lemma 2.17]{FB-CTDS}, there exists a nonsingular matrix $P_c$ such that $w(P_c A_c P_c^{-1})<1$. Consequently, one can show that
\begin{align*}
&\|x(k) - x^\star\|_P \leq 2 \left[w(P_c A_c P_c^{-1})\right]^k \|x(0) - x^\star\|_P,
\end{align*}
where $P = \mathrm{diag}\{I, P_c\}V^*$ and the weighted norm is defined as $\Vert{}x\Vert_P := \Vert P x\Vert$.
\end{remark}

\begin{remark}
    The gap between the numerical-radius convergence factor and the asymptotic convergence rate governed by $\rho(A)$ is related to the degree of non-normality of $A$, as in Lemma~\ref{lemma:depature_from_normality}. For normal $A$, $w(A)=\rho(A)$.
    %and the numerical-radius error bound \eqref{eq: w-contractive_error_bound} has the exact asymptotic decay factor up to a multiplicative constant $2$. 
    For highly nonnormal matrices, the gap may become large. The advantages of numerical-radius contraction are two folds. First, it provides a finite-step error bound \eqref{eq: w-contractive_error_bound}, whereas the spectral radius only characterizes the asymptotic convergence rate. Second, it gives an explicit robustness margin against perturbations.
    %which may not be inferred from the spectral radius. 
    Specifically, whenever $w(A)<1$, the inequality $w(A+\Delta)\leq w(A)+\|\Delta\|$ guarantees convergence for all perturbations satisfying $\|\Delta\|<1-w(A)$. We mention that developing sharper convergence bounds under numerical-radius contraction remains an interesting direction for future research.
\end{remark}

\subsection{Iterative methods for equations with sectorial matrices}
Consider the system of linear equations
\begin{align}
    Tx=b,\label{secequation}
\end{align}
where $T\!\in\!\mathbb{C}^{n\times n}$ is (semi-)sectorial, and $b\!\in\!\mathbb{C}^n$. We shall establish iterative methods for solving this class of equations by utilizing the concept of numerical-radius contraction. This prepares the later development of iterative methods for solving feedback equations under small phase condition.

\begin{figure}[htbp]
\centering
\subfloat[$T$ is sectorial]{%
  \begin{minipage}{0.45\linewidth}
    \centering
    \includegraphics[width=1\linewidth]{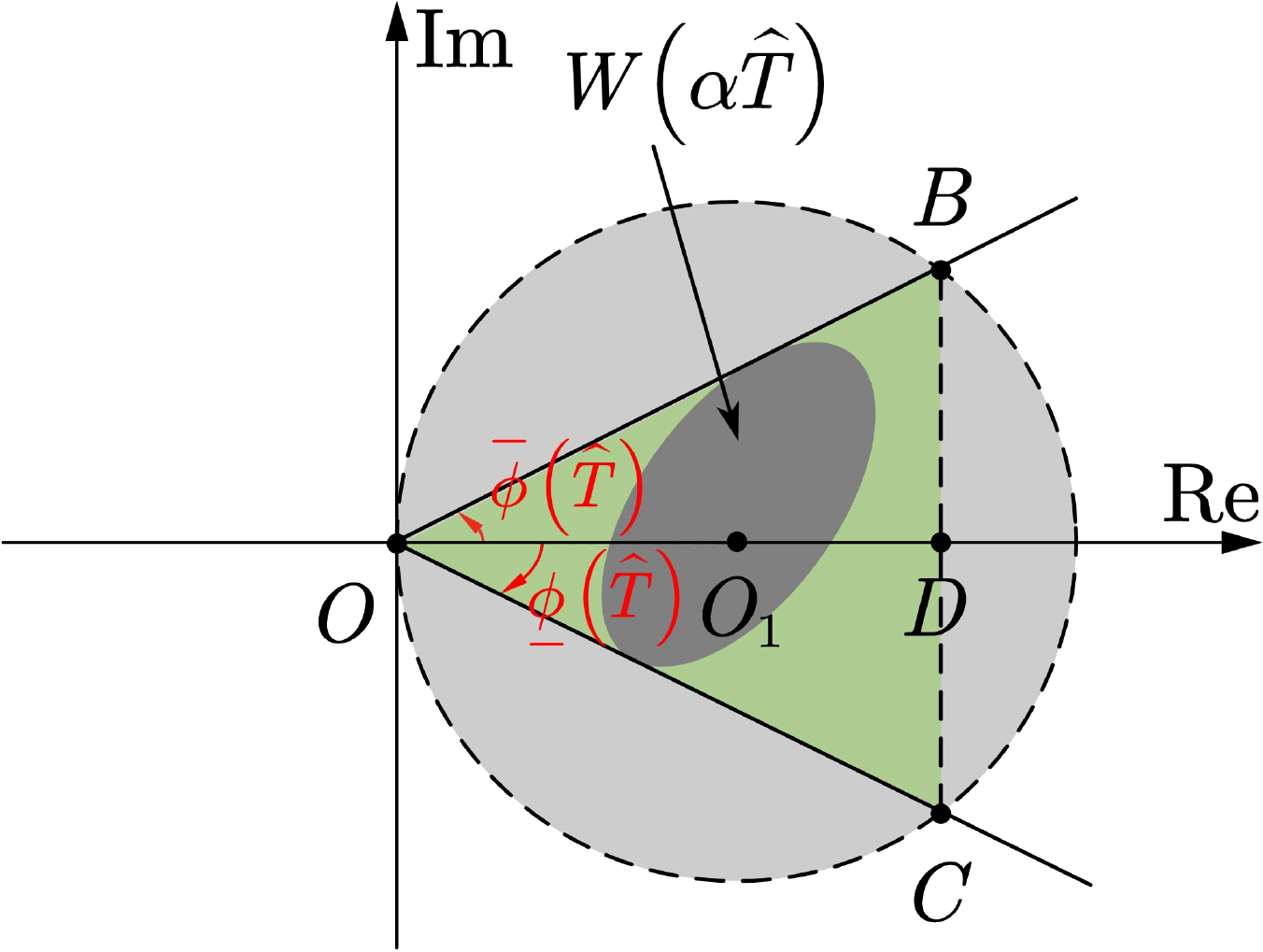}
  \end{minipage}
  \label{fig: Graphical_Sectorial}
}
\hfil
\subfloat[$T$ is quasi-sectorial]{%
    \begin{minipage}{0.45\linewidth}
    \centering
    \includegraphics[width=1\linewidth]{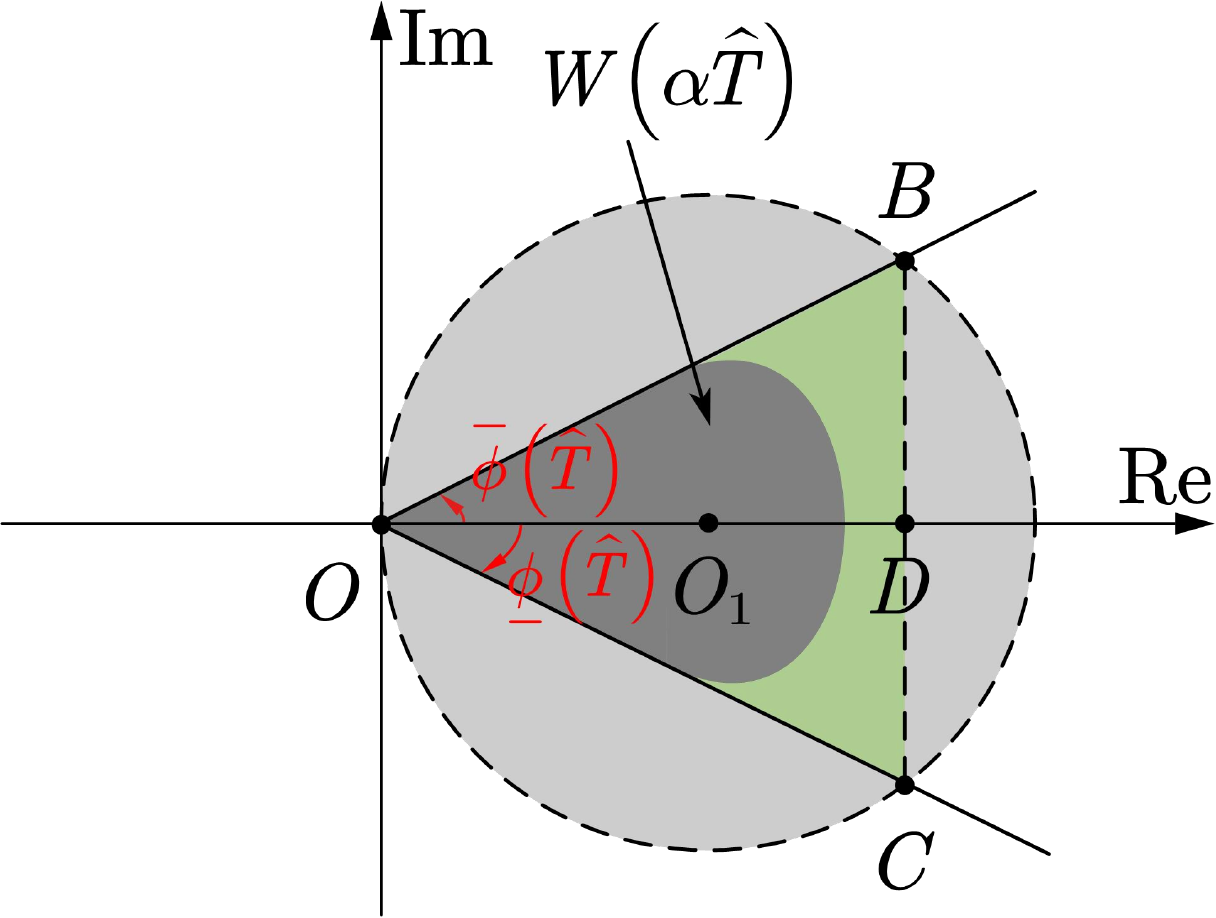}
  \end{minipage}
  \label{fig: Graphical_Quasi}
}
\caption{Geometric interpretation of the core idea underlying the iterative method~\eqref{eq: sectorial-iteration}. Here, $O$ is the origin, and $O_1$ is the point $1$. The region in light grey is an open unit disk centered at $1$. A sector anchored at the origin, which is symmetric with respect to the real axis with opening angle $\delta(T)$, intersects the unit circle, forming a triangle $\Delta OBC$. The line segment $BC$ is perpendicular to the real axis and intersects the real axis at point $D$. 
The region in dark grey represents $W(\alpha\hat{T})$.
}
\label{fig: Graphical_Interpre}
\end{figure}

\begin{lemma}\label{lemma: sectorial iteration}
    Let $T\!\in\!\mathbb{C}^{n\times n}$ be sectorial. Then the equation (\ref{secequation}) has a unique solution $x^\star$. Moreover, for any $0<\alpha<\displaystyle\frac{1+\cos\delta(T)}{\mu(e^{-j\gamma(T)}T)}
    $, the iteration
    \begin{equation}\label{eq: sectorial-iteration}
        x(k+1)=(I-\alpha e^{-j\gamma(T)} T)x(k)+\alpha e^{-j\gamma(T)} b
    \end{equation}
    converges to $x^\star$ for any $x(0)\in\mathbb{C}^n$ with an error bound
    \begin{equation}
        \Vert x(k)-x^\star \Vert\leq 2[w(I-\alpha e^{-j\gamma(T)} T)]^k\Vert x(0)-x^\star \Vert.
    \end{equation}
\end{lemma}
\begin{pf}
    Since $T$ is sectorial, it is nonsingular, ensuring the existence and uniqueness of the solution $x^\star$. 
    
    Let $\widehat{T}=e^{-j\gamma(T)}T$ and $\widehat{b}=e^{-j\gamma(T)}b$.  The iteration (\ref{eq: sectorial-iteration}) can be rewritten as
    \begin{equation}\label{eq: accretive_iteration}
        x(k+1)=(I-\alpha \widehat{T})x(k)+\alpha \widehat{b}.
    \end{equation}
    Clearly, such iteration has a unique fixed point identical to the solution $x^\star$ of the equation $Tx=b$.

    We now establish convergence of iteration~\eqref{eq: accretive_iteration}. Note that $\widehat{T}$ is strictly accretive with phases $\overline{\phi}(\widehat{T})=\frac{\delta(T)}{2}\in [0,\frac{\pi}{2})$ and $\underline{\phi}(\widehat{T})=-\frac{\delta(T)}{2}$. Consider a geometric illustration as shown in Fig. \ref{fig: Graphical_Interpre}(a). A direct calculation yields $\overline{OD}=1+\cos(2\overline{\phi}(\widehat{T}))=1+\cos(\delta(T))$. Hence, for any  $0<\alpha<\displaystyle\frac{1+\cos(\delta(T))}{\mu(e^{-j\gamma(T)}T)}$, we have $\mu(\alpha \widehat{T})=\alpha \mu(\hat{T})<\overline{OD}$. Consequently, the numerical range $W(\alpha\widehat{T})$ lies strictly inside the triangle $\Delta OBC$, and in turn lies inside the open unit disk centered at $1$. This implies that $W(I-\alpha\widehat{T})$ lies inside an open unit circle centered at the origin and, thus, $I-\alpha\widehat{T}$ is $w$-contractive. The convergence of iteration~\eqref{eq: accretive_iteration}, along with the stated error bound, then follows directly from Lemma~\ref{lemma: NR_contractive_convergence}(i).
\end{pf}

We mention that the quantities involved in the step size bound in Lemma \ref{lemma: sectorial iteration} can be efficiently computed. Specifically, $\gamma(T),\delta(T)$ can be computed via LMI solvers \cite{wang2023phases}, and $\mu(e^{-j\gamma(T)}T)$ can be computed using standard eigensolvers.

The iterative method as shown in Lemma \ref{lemma: sectorial iteration} is closely related to the forward step method for linear equations involving a strictly accretive $T$ \cite{ryu2022large}, but with an important distinction. The forward step method yields an iteration
\begin{equation}\label{eq: forward-step iteration}
x(k+1)=(I-\alpha T)x(k)+\alpha b,
\end{equation}
where $\alpha>0$ is chosen to make the matrix $I-\alpha T$ contractive in the conventional sense, i.e., has norm less than $1$. In contrast, in the proposed iteration~\eqref{eq: sectorial-iteration}, $\alpha>0$ is chosen so that the matrix $I-\alpha e^{-j\gamma(T)} T$ is numerical radius contractive, a less conservative condition than norm contraction. Phase information is exploited to make $e^{-j\gamma(T)}T$ a strictly accretive matrix and to derive a sufficient step-size condition. Moreover, the numerical-radius analysis provides an error bound whose geometric factor can be smaller than that obtained from the norm of the iteration matrix.

\begin{remark}
When $\delta(T)=0$, the rotated matrix $e^{-j\gamma(T)}T$ is positive definite,
and the step-size bound in Lemma~\ref{lemma: sectorial iteration} reduces to $2/\lambda_{\max}(e^{-j\gamma(T)}T)$, coinciding with the classical gradient-descent bound \cite{ryu2022large}.
\end{remark}

The following lemma shows that if $T$ is relaxed to be quasi-sectorial, the iteration~\eqref{eq: sectorial-iteration} still converges, but the limit point is dependent on the initial condition.
\begin{lemma}\label{lemma: quasi_sectorial iteration}
   Let $T\!\in\!\mathbb{C}^{n\times n}$ be quasi-sectorial and $b\!\in\!\mathcal{R}(T)$. Then for any $0\!<\!\alpha\!<\!\displaystyle \frac{1+\cos\delta(T)}{\mu(e^{-j\gamma(T)}T)}$, the iteration~\eqref{eq: sectorial-iteration} converges to a solution 
   \[x^\star=V_1V_1^*x(0)+T^{\dagger}b,\] 
  where $V_1$ is an isometry whose columns form a basis of $\mathcal{N}(T)$. Moreover, the following error bound holds:
    \[
      \Vert x(k)-x^\star \Vert \leq 2[w(I-\alpha e^{-j\gamma(T)}T_c) ]^k \Vert x(0)-x^\star \Vert,
    \]
    where $T_c={V_1^\perp}^*TV_1^\perp$ with $V_1^\perp$ being an orthogonal complement of $V_1$.
\end{lemma}
\begin{pf}
    Using geometric configuration as in Fig.~\ref{fig: Graphical_Interpre}(b) and arguments in analogy to those in the proof of Lemma~\ref{lemma: sectorial iteration}, one can show that for $0\!<\!\alpha\!<\!\displaystyle \frac{1+\cos(\delta(T))}{\mu(e^{-j\gamma(T)}T)}$, the matrix $I-\alpha e^{-j\gamma(T)}T$ is $w$-semi-contractive, with $1$ being a sharp point of its numerical range. The convergence of iteration~\eqref{eq: sectorial-iteration} and the corresponding error bound then follow directly from Lemma~\ref{lemma: NR_contractive_convergence}(ii).
\end{pf}

 Things become more complicated when $T$ is further relaxed to be semi-sectorial. It can be inferred from Lemma~\ref{lemma: NR_contractive_convergence}(ii) that the iteration~\eqref{eq: sectorial-iteration} may converge for a sufficiently small $\alpha>0$ if $e^{-j\gamma(T)}T$ has no imaginary-axis eigenvalues except possibly at $0$. Nevertheless, the iteration~\eqref{eq: sectorial-iteration} may fail to converge for a semi-sectorial $T$ in general. For instance, let $T\!=\!\begin{bmatrix}
    0 & 1 \\
    -1 & 0
\end{bmatrix}$. Its numerical range is $W(T)=\left\{z=j\omega:\omega\in[-1,1] \right\}$. Thus, $T$ is semi-sectorial with phases $\overline{\phi}(T)=-\underline{\phi}(T)=\frac{\pi}{2}$. Note that $\rho(I-\alpha \widehat{T})=\sqrt{1+\alpha^2}\geq 1$ for any $\alpha\in \mathbb{R}$. Hence, the iteration (\ref{eq: sectorial-iteration}) will diverge away from the unique solution $\begin{bmatrix}
    x_1^\star \\
    x_2^\star
\end{bmatrix}=\begin{bmatrix}
    -b_2 \\
    b_1
\end{bmatrix}$ unless it starts at the solution.

\section{Iterative Methods for Feedback Equations under Small Phase Condition}\label{section: linear equation case}
In this section, we shall develop a novel iterative method for solving the feedback equation (\ref{eq: core_feedback_equation}) under the small phase condition. Iterative methods based on the small gain condition have been addressed in the literature, and one representative algorithm is briefly revisited below.

Let $G, H\in\mathbb{C}^{n\times n}$ satisfy the small gain condition, namely, $\Vert G\Vert \cdot \Vert H\Vert<1$. By Lemma \ref{lemma: small gain condition}, the feedback equation (\ref{eq: core_feedback_equation}) has a unique solution. To construct a contractive fixed-point iteration for (\ref{eq: core_feedback_equation}), we introduce a scaling parameter $\gamma>0$, chosen as follows. If $G=0$, choose any $\gamma>\|H\|$. Otherwise, choose $\gamma$ so that $\|H\|<\gamma<\|G \|^{-1}$. In either case, the scaled matrices $\overline{G}=\gamma G$ and $\overline{H}=\gamma^{-1}H$ both have norms strictly less than $1$. Accordingly, the feedback-type equation \eqref{eq: core_feedback_equation} can be rewritten as
\begin{equation}\label{eq: small_gain_feedback_rewritten}
    \begin{bmatrix}
    I & \overline{H} \\
    -\overline{G} & I
    \end{bmatrix}\begin{bmatrix}
        \overline{u}_1 \\
        \overline{u}_2
    \end{bmatrix}=\begin{bmatrix}
        \overline{w}_1 \\
        \overline{w}_2
    \end{bmatrix}
\end{equation}
where $\overline{u}_1=u_1$, $\overline{u}_2=\gamma u_2$, $\overline{w}_1=w_1$ and $\overline{w}_2=\gamma w_2$. To compute the solution iteratively, note that \begin{align*}
    &\begin{bmatrix}
		I & \overline{H} \\
		-\overline{G} & I
	\end{bmatrix}\begin{bmatrix}
		\overline{u}_1\\
		\overline{u}_2
	\end{bmatrix}=\begin{bmatrix}
		\overline{w}_1\\
		\overline{w}_2
	\end{bmatrix}  \\
    &\Leftrightarrow \left(\begin{bmatrix}
    I & 0 \\
    0 & I
\end{bmatrix}-\begin{bmatrix}
    0 & -\overline{H} \\
    \overline{G} & 0
\end{bmatrix}\right)\begin{bmatrix}
	      \overline{u}_1\\
		\overline{u}_2
	\end{bmatrix}=\begin{bmatrix}
		\overline{w}_1\\
		\overline{w}_2
	\end{bmatrix} \\
    & \Leftrightarrow \begin{bmatrix}
		\overline{u}_1\\
		\overline{u}_2
	\end{bmatrix}=\begin{bmatrix}
    0 & -\overline{H} \\
    \overline{G} & 0
\end{bmatrix}\begin{bmatrix}
		\overline{u}_1\\
		\overline{u}_2
	\end{bmatrix}+\begin{bmatrix}
		\overline{w}_1\\
		\overline{w}_2
	\end{bmatrix}.
\end{align*}
This naturally leads to the fixed-point iteration
    \begin{equation*}
        \begin{bmatrix}
            \overline{u}_1(k+1) \\
            \overline{u}_2(k+1)
        \end{bmatrix}=\begin{bmatrix}
            0 & -\overline{H} \\
            \overline{G} & 0
        \end{bmatrix}\begin{bmatrix}
            \overline{u}_1(k)\\
            \overline{u}_2(k)
        \end{bmatrix}+\begin{bmatrix}
            \overline{w}_1 \\
            \overline{w}_2
        \end{bmatrix}.
    \end{equation*}
Since $\left\|\begin{bmatrix} 0 & -\overline{H} \\ \overline{G}  & 0\end{bmatrix}\right\| = \max\{\|\overline{G} \|, \| \overline{H} \|\}\!<\!1$,
the above iteration converges to the unique solution of (\ref{eq: small_gain_feedback_rewritten}) for any $\begin{bmatrix}
            \overline{u}_1(0) \\
            \overline{u}_2(0)
        \end{bmatrix}\in\mathbb{C}^{2n}$. Finally, the solution of feedback equation \eqref{eq: core_feedback_equation} can be recovered as $u_1=\overline{u}_1$ and $u_2=\gamma^{-1}\overline{u}_2$.

Now we consider the situation where $G,H$ satisfy the small phase condition as in Lemma \ref{lemma:small_phase_condition}. Clearly, when the condition holds, at least one of $G$ and $H$ is quasi-sectorial, i.e., having field angle less than $\pi$. Without loss of generality, we assume $G$ is quasi-sectorial and $H$ is semi-sectorial.

We first examine the special case where both $G$ and $H$ are sectorial. In this case, the small phase condition guarantees that $W(G)$ and $W(H)$ can be rotated by equal angles in opposite directions so that both numerical ranges lie in the open right half-plane after rotation, i.e., there exists a real number $\phi_0$ satisfying
\begin{equation*}
\begin{cases}
    \phi_0< \min \left\{ -\overline{\phi}(G)+\frac{\pi}{2},\underline{\phi}(H)+\frac{\pi}{2} \right\} ,\\
    \phi_0> \max \left\{ -\underline{\phi}(G)-\frac{\pi}{2},\overline{\phi}(H)-\frac{\pi}{2} \right\},
\end{cases}
\end{equation*}
so that $\widehat{G}\!=\!e^{j\phi_0}G$ and $\widehat{H}\!=\!e^{-j\phi_0}H$ are both strictly accretive. Consequently, the feedback-type equation (\ref{eq: core_feedback_equation}) can be rewritten as 
\begin{equation}\label{eq: gain_rewritten_feedback_accretive}
    \begin{bmatrix}
        I & \widehat{H} \\
        -\widehat{G} & I
    \end{bmatrix}\begin{bmatrix}
        \widehat{u}_1 \\
        \widehat{u}_2
    \end{bmatrix}=\begin{bmatrix}
        \widehat{w}_1 \\
        \widehat{w}_2
    \end{bmatrix},
\end{equation}
where $\widehat{u}_1=u_1, \widehat{u}_2=e^{j\phi_0}u_2,\widehat{w}_1=w_1,$ and $\widehat{w}_2=e^{j\phi_0}w_2$. Multiplying both sides of (\ref{eq: gain_rewritten_feedback_accretive}) by $\begin{bmatrix}
    0 & -I \\
    I & 0
\end{bmatrix}$ yields
\begin{equation}\label{eq: phase_rewritten_feedback_accretive}
    \begin{bmatrix}
        \widehat{G} & -I \\
        I & \widehat{H}
    \end{bmatrix}\begin{bmatrix}
        \widehat{u}_1 \\
        \widehat{u}_2
    \end{bmatrix}=\begin{bmatrix}
        -\widehat{w}_2 \\
        \widehat{w}_1
    \end{bmatrix}.
\end{equation}
Clearly, $\begin{bmatrix}
    \widehat{G} & -I \\
        I & \widehat{H}
\end{bmatrix}$ is strictly accretive since
\begin{equation*}
    \mathrm{Re}\left( \begin{bmatrix}
        \widehat{G} & -I \\
        I & \widehat{H}
    \end{bmatrix}\right)=\begin{bmatrix}
        \mathrm{Re}(\widehat{G}) & 0 \\
        0 & \mathrm{Re}(\widehat{H})
    \end{bmatrix}>0.
\end{equation*}
Then, equation (\ref{eq: phase_rewritten_feedback_accretive}) can be solved using the iterative method previously established in Lemma \ref{lemma: sectorial iteration}. Finally, the solution of feedback equation (\ref{eq: core_feedback_equation}) can be recovered as $\begin{bmatrix}u_1\\u_2\end{bmatrix} =\begin{bmatrix}\widehat{u}_1\\e^{-j\phi_0}\widehat{u}_2\end{bmatrix}$.

Things become more complicated if $G$ remains sectorial but $H$ is relaxed to be only semi-sectorial. In this case, one can find $\phi_0$ satisfying 
\begin{equation}\label{eq: phi0_select_condition}
\begin{cases}
    \phi_0\leq \min \left\{ -\overline{\phi}(G)+\frac{\pi}{2},\underline{\phi}(H)+\frac{\pi}{2} \right\},\\
    \phi_0\geq \max \left\{ -\underline{\phi}(G)-\frac{\pi}{2},\overline{\phi}(H)-\frac{\pi}{2} \right\},
\end{cases}
\end{equation}
such that $\widehat{G}\!=\!e^{j\phi_0}G$ is strictly accretive and $\widehat{H}\!=\!e^{-j\phi_0}H$ is accretive. Then $\begin{bmatrix}
    \widehat{G} & -I \\
        I & \widehat{H}
\end{bmatrix}$ is accretive but not strictly accretive in general, and thus the iterative method in Lemma \ref{lemma: sectorial iteration} cannot be directly applied. To address this issue, we consider a convex combination of equations \eqref{eq: gain_rewritten_feedback_accretive} and \eqref{eq: phase_rewritten_feedback_accretive}, hoping to exploit the identity blocks on the diagonal of \eqref{eq: gain_rewritten_feedback_accretive} to make the resulting coefficient matrix strictly accretive. The following theorem establishes the validity of this approach.

\begin{theorem}\label{Thm: passive_iteration}
   Let $G$ be sectorial and $H$ be semi-sectorial such that the small phase condition (\ref{eq: matrix_small_phase_condition}) holds. Let $\phi_0$ be chosen as in (\ref{eq: phi0_select_condition}) such that $\widehat{G}=e^{j\phi_0}G$ is strictly accretive and $\widehat{H}=e^{-j\phi_0}H$ is accretive.
Then the following statements hold. 
    \begin{enumerate}[(i)]
    \item The matrix
    \begin{equation}\label{Ttheta}
        T_{\theta}=\theta \begin{bmatrix}
            I & \widehat{H} \\
            -\widehat{G} & I
        \end{bmatrix}+(1-\theta)\begin{bmatrix}
            \widehat{G} & -I \\
            I & \widehat{H}
        \end{bmatrix}
    \end{equation}
    is strictly accretive for any $0<\theta<\theta_0$, where 
\begin{align}\label{theta0}
\theta_0 \!=\!
\begin{cases}
1,  \text{ if } \left\Vert \dfrac{\widehat{H} - \widehat{G}^*}{2} \right\Vert \leq 1, \\[12pt]
\!\dfrac{\lambda_{\min}(\mathrm{Re}(\widehat{G})) + \lambda_{\min}(\mathrm{Re}(\widehat{H}))}
{\lambda_{\min}(\mathrm{Re}(\widehat{G})) \!+\! \lambda_{\min}(\mathrm{Re}(\widehat{H})) \!+\! \left\Vert \dfrac{\widehat{H} - \widehat{G}^*}{2} \right\Vert^2 \!-\! 1},
 \\\hspace{160pt}\text{otherwise.}
\end{cases}
\end{align}
\item 
    For any  $0<\alpha<\displaystyle\frac{1+\cos\delta(T_\theta)}{\mu( e^{-j\gamma(T_\theta)} T_\theta)}$ and $0<\theta<\theta_0$, the iteration
    \begin{multline}
        \begin{bmatrix}
            \widehat{u}_1(k+1)\\
            \widehat{u}_2(k+1)
        \end{bmatrix}=(I-\alpha e^{-j\gamma(T_\theta)}T_\theta)\begin{bmatrix}
            \widehat{u}_1(k)\\
            \widehat{u}_2(k)
        \end{bmatrix}\\
        +\alpha e^{-j\gamma(T_\theta)}\begin{bmatrix}
            \theta \widehat{w}_1-(1-\theta)\widehat{w}_2 \\
            (1-\theta)\widehat{w}_1+\theta \widehat{w}_2
        \end{bmatrix}\label{eq: passive_iteration}
    \end{multline}
    will converge to the unique solution  $\begin{bmatrix}
        \widehat{u}_1^\star \\
        \widehat{u}_2^\star
    \end{bmatrix}$ of (\ref{eq: gain_rewritten_feedback_accretive}) for any initial condition with an error bound    
    \begin{multline*}
        \left\Vert \begin{bmatrix}
            \widehat{u}_1(k)\\
            \widehat{u}_2(k)
        \end{bmatrix}-\begin{bmatrix}
            \widehat{u}_1^\star \\
            \widehat{u}_2^\star
        \end{bmatrix}\right\Vert \\
        \leq 2[w(I-\alpha e^{-j\gamma(T_\theta)}T_\theta)]^k\left\Vert \begin{bmatrix}
            \widehat{u}_1(0)\\
            \widehat{u}_2(0)
        \end{bmatrix}-\begin{bmatrix}
            \widehat{u}_1^\star \\
            \widehat{u}_2^\star
        \end{bmatrix}\right\Vert.
    \end{multline*}
    Moreover, the solution of equation (\ref{eq: core_feedback_equation}) can be recovered as $\begin{bmatrix}u_1\\u_2\end{bmatrix} =\begin{bmatrix}\widehat{u}_1^\star\\e^{-j\phi_0}\widehat{u}_2^\star\end{bmatrix}$.
    \end{enumerate}
\end{theorem}

\begin{pf}
    See Appendix \ref{appdix_proof: gain_phase_convex}.
\end{pf}

% Based on Theorem \ref{Thm: passive_iteration}, we can construct an iterative method for solving the feedback-type equation in the general case where $G$ is quasi-sectorial and $H$ is semi-sectorial. By the small phase condition~\eqref{eq: matrix_small_phase_condition}, 
For the general case where $G$ is a quasi-sectorial matrix and $H$ is a semi-sectorial matrix, one can choose $\phi_0$ as in (\ref{eq: phi0_select_condition})
such that $\widehat{G}\!=\!e^{j\phi_0}G$ is quasi-strictly accretive and $\widehat{H}\!=\!e^{-j\phi_0}H$ is accretive. The feedback equation (\ref{eq: core_feedback_equation}) can then be rewritten in the form (\ref{eq: gain_rewritten_feedback_accretive}). Since $\widehat{G}$ is quasi-strictly accretive, it admits the decomposition 
\begin{align}
    U^*\widehat{G}U&=\begin{bmatrix}
        0 & 0 \\
        0 & \widetilde{G}_{22}
    \end{bmatrix},\label{dechatG}
\end{align}
where $U$ is unitary and $\widetilde{G}_{22}$ is strictly accretive. Accordingly, applying a congruence transformation on $\widehat{H}$ with $U$ yields
\begin{align}
    U^*\widehat{H}U=\begin{bmatrix}
        \widetilde{H}_{11} & \widetilde{H}_{12} \\
        \widetilde{H}_{21} & \widetilde{H}_{22}
    \end{bmatrix}, \label{eq: second_decomposition}
\end{align}
where $\widetilde{H}_{22}$ is accretive. Now, multiplying both sides of (\ref{eq: gain_rewritten_feedback_accretive}) by $\begin{bmatrix}
    U^* & 0\\
    0 & U^*
\end{bmatrix}$, we have
\begin{multline*}
    \begin{bmatrix}
        I & U^*\widehat{H}U \\
        -U^*\widehat{G}U & I
    \end{bmatrix}\begin{bmatrix}
        U^*\widehat{u}_1 \\
        U^* \widehat{u}_2
    \end{bmatrix}=\begin{bmatrix}
        U^*\widehat{w}_1 \\
        U^* \widehat{w}_2
    \end{bmatrix} \\
    \Leftrightarrow     \begin{bmatrix}
        I & 0 & \widetilde{H}_{11} & \widetilde{H}_{12} \\
        0 & I & \widetilde{H}_{21} & \widetilde{H}_{22} \\
        0 & 0 & I & 0 \\
        0& -\widetilde{G}_{22} & 0 & I
    \end{bmatrix}\begin{bmatrix}
        \widetilde{u}_{11} \\
        \widetilde{u}_{12} \\
        \widetilde{u}_{21} \\
        \widetilde{u}_{22}
    \end{bmatrix}=\begin{bmatrix}
        \widetilde{w}_{11} \\
        \widetilde{w}_{12} \\
        \widetilde{w}_{21} \\
        \widetilde{w}_{22}
    \end{bmatrix},
\end{multline*}
where $\widetilde{u}_{ij}=U_j^*\widehat{u}_i$ and $\widetilde{w}_{ij}=U_j^*\widehat{w}_i$, $i,j=1,2$. Clearly, $\widetilde{u}_{21}=\widetilde{w}_{21}$. The rest of the above equation can be partitioned into two parts:
\begin{align}
        &\widetilde{u}_{11}=\widetilde{w}_{11}-\widetilde{H}_{11}\widetilde{w}_{21}-\widetilde{H}_{12}\widetilde{u}_{22},
\label{eq: first_part_equation} \\
    &\hspace{-3pt}\begin{bmatrix}
        I & \widetilde{H}_{22} \\
        -\widetilde{G}_{22} & I
    \end{bmatrix}\begin{bmatrix}
        \widetilde{u}_{12} \\
        \widetilde{u}_{22}
    \end{bmatrix}=\begin{bmatrix}
        \widetilde{w}_{12}-\widetilde{H}_{21}\widetilde{w}_{21} \label{eq: second_part_equation}\\
        \widetilde{w}_{22}
    \end{bmatrix}.
\end{align}
Recall that $\widetilde{G}_{22}$ is strictly accretive and $\widetilde{H}_{22}$ is accretive. The solution of (\ref{eq: second_part_equation}), i.e., $\widetilde{u}_{12}$ and $\widetilde{u}_{22}$, can be obtained by the iterative method described in Theorem \ref{Thm: passive_iteration}. Then $\widetilde{u}_{11}$ can be obtained by substituting $\widetilde{u}_{22}$ into (\ref{eq: first_part_equation}). Finally, the solution of equation (\ref{eq: core_feedback_equation}) can be recovered as
\begin{equation*}
    u_1=U\begin{bmatrix}
        \widetilde{u}_{11} \\
        \widetilde{u}_{12}
    \end{bmatrix},u_2=e^{-j\phi_0}U\begin{bmatrix}
        \widetilde{u}_{21} \\
        \widetilde{u}_{22}
    \end{bmatrix}.
\end{equation*}

\begin{remark}
A numerical step in the above procedure is to construct the decomposition (\ref{dechatG}). Let $U=\begin{bmatrix}U_1&U_2\end{bmatrix}$, where the columns of $U_1$ form an orthonormal basis of $\mathcal{N}(\widehat{G})$. Since $\widehat{G}$ is quasi-strictly accretive, it follows from Lemma \ref{lemma: when_semi_to_quasi} that
$\mathcal{N}(\widehat{G}) = \mathcal{N}(\mathrm{Re}(\widehat{G}))$, and the computational cost scales approximately as $O(\mathrm{nnz}(\operatorname{Re}(\widehat G))\cdot \dim\mathcal N(\widehat G))$ \cite{golubvanloan2013}.
\end{remark}

The proposed method for solving the feedback-type equation (1) is summarized in Algorithm \ref{alg:proposed_method}.

\begin{algorithm}[htbp]
\caption{ Iterative Solver for Feedback-Type Linear Equations under Small Phase Condition}
\label{alg:proposed_method}
\begin{algorithmic}[1]
\REQUIRE $G,H\in\mathbb{C}^{n\times n}$, $w_1,w_2\in\mathbb{C}^n$,
tolerance $\varepsilon>0$, maximum iteration number $K_{\max}$.
\ENSURE The solution $(u_1,u_2)$ of \eqref{eq: core_feedback_equation}.

\STATE Verify the small phase condition \eqref{eq: matrix_small_phase_condition}; terminate if it is not satisfied.
\STATE Choose $\phi_0$ according to \eqref{eq: phi0_select_condition} and set
$\widehat G=e^{j\phi_0}G$, $\widehat H=e^{-j\phi_0}H$.
\STATE If $\operatorname{Re}(\widehat G)$ is singular, construct
$U=[U_1\ U_2]$ and apply the decompositions \eqref{eq: first_part_equation}--\eqref{eq: second_part_equation}.
\STATE Choose $0<\theta<\theta_0$ according to Theorem~\ref{Thm: passive_iteration}, construct $T_\theta$ in
\eqref{Ttheta}, and select $0<\alpha<
\frac{1+\cos\delta(T_\theta)}
{\mu(e^{-j\gamma(T_\theta)}T_\theta)}.$
\STATE Set $b_\theta=\bigl[\begin{smallmatrix}
\theta\widehat w_1-(1-\theta)\widehat w_2\\
(1-\theta)\widehat w_1+\theta\widehat w_2
\end{smallmatrix}\bigr]$ and initialize $(\widehat{u}_1(0),\widehat{u}_2(0))$.
\FOR{$k=0,\ldots,K_{\max}-1$}
    \STATE Update $(\widehat{u}_1(k+1),\widehat{u}_2(k+1))$ according to \eqref{eq: passive_iteration}.
    \IF{$\left\|b_\theta-T_\theta \bigl[\begin{smallmatrix}
\widehat{u}_1(k+1)\\
\widehat{u}_2(k+1)
\end{smallmatrix}\bigr]\right\|/\|b_\theta\|\leq\varepsilon$}
        \STATE \textbf{break}.
    \ENDIF
\ENDFOR
\STATE Recover $(u_1,u_2)$ by reversing the nullspace reduction (if applied) and phase rotation.
\RETURN $(u_1,u_2)$.
\end{algorithmic}
\end{algorithm}

\begin{remark}
How to select parameters $\phi_0$, $\theta$, and $\alpha$ to achieve fastest convergence is generally challenging. Some preliminary heuristic strategies for parameter selection are discussed in Appendix~\ref{appdix: parameter_selection}.
\end{remark}

\section{Case Study}\label{section: example}
\subsection{Power-network Model}
\begin{figure}[htbp]
\centering
    \includegraphics[width=0.7\linewidth]{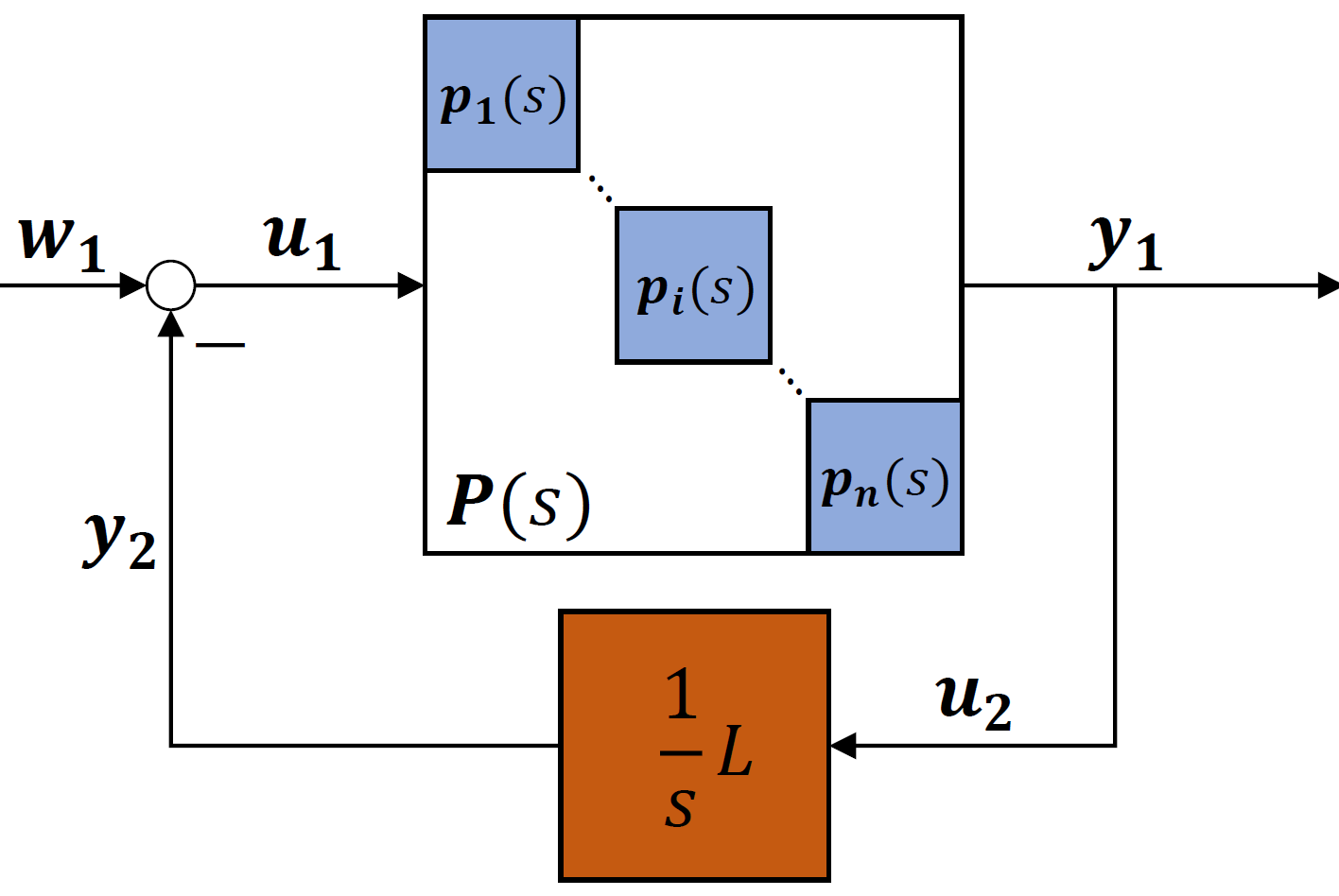}
  \caption{An $n$-bus power network.}
   \label{fig: block_diagram}
\end{figure}

Consider an $n$-bus power network modeled as a feedback interconnection of bus dynamics $\boldsymbol{P}(s)\!=\!\mathrm{diag}\{p_1(s),\dots,p_n(s)\}$ and network dynamics $\frac{1}{s}L$ as shown in Fig. \ref{fig: block_diagram} \cite{Pates2019,Paganini2020,Jiang2021}. The bus dynamics $p_i(s)$ is a scalar closed-loop transfer function representing a local feedback interconnection of a generator $g_i(s)$ and inverter $c_i(s)$, i.e., $p_i(s)=\frac{g_i(s)}{1-g_i(s)c_i(s)}$. The network dynamics is given by $\boldsymbol{y_2}(s)=\frac{1}{s}L\boldsymbol{u_2}(s)$, where $L=[l_{ij}]\in \mathbb{R}^{n\times n}$ is a weighted Laplacian matrix with entries given by
\[
l_{ij}=\left. \frac{\partial}{\partial \theta_j}\sum_{l=1}^n |V_i||V_l|b_{il}\sin(\theta_i-\theta_l) \right|_{\theta=\theta_0}.
\]
Here, $\theta_i$ represents the voltage angle at bus $i$ with $\theta_0$ being the equilibrium angles, $|V_i|$ is the voltage magnitude at bus $i$, and $b_{il}\geq 0$ represents the susceptance of the transmission line $i\to l$ ($b_{il}=0$ if there is no line). 

The closed-loop dynamics can be written as
\begin{equation}\label{eq: power_network_equation}
\begin{bmatrix}
    I & \frac{1}{s}L \\
    -\boldsymbol{P}(s) & I
\end{bmatrix}
\begin{bmatrix}
    \boldsymbol{u_1}(s) \\
    \boldsymbol{u_2}(s)
\end{bmatrix}
=
\begin{bmatrix}
    \boldsymbol{w_1}(s) \\
    0
\end{bmatrix},
\end{equation}
where $\boldsymbol{w_1}$, $\boldsymbol{u_1}$, $\boldsymbol{u_2}=\boldsymbol{y_1}$, and $\boldsymbol{y_2}$ denote the power disturbance, the net power imbalance, the bus frequency deviation, and the network power fluctuation, respectively.

For each nonzero frequency $s=j\omega$, \eqref{eq: power_network_equation} reduces to a feedback-type linear equation
\[
\begin{bmatrix}
I & \frac{1}{j\omega}L \\
-\boldsymbol{P}(j\omega) & I
\end{bmatrix}
\begin{bmatrix}
\boldsymbol{u_1}(j\omega)\\
\boldsymbol{u_2}(j\omega)
\end{bmatrix}
=
\begin{bmatrix}
\boldsymbol{w_1}(j\omega)\\
0
\end{bmatrix}.
\]
Evaluating the network frequency response amounts to solving such feedback-type linear equations parameterized by $\omega$. %\sout{Note that such feedback interconnections do not satisfy the small-gain condition, as the term $\frac{1}{j\omega}L$ exhibits large gain in low frequency. On the contrary, the small-phase condition is more natural and suited in this context.} 

Note that in power networks, the matrices involved are typically sparse: $\boldsymbol{P}(j\omega)$ is diagonal and $L$ contains nonzero off-diagonal entries only for physically connected buses. Direct methods for solving the resulting feedback equations typically require $\mathcal{O}(n^3)$ operations. In contrast, the proposed iterative method only involves matrix–vector multiplications with complexity $\mathcal{O}(k\cdot\mathrm{nnz}(T))$, where $\mathrm{nnz}(T)$ is the number of nonzero entries in the iteration matrix $T$, and $k$ is the number of iterations. For dense matrices $\mathrm{nnz}(T)\approx n^2$. Nevertheless, practical power networks are often highly sparse. As an illustrative example, the IEEE 300-bus test system contains only 304 transmission lines. In such cases, $\mathrm{nnz}(T)$ scales with the number of transmission lines rather than $n^2$, giving the proposed iterative method a significant computational advantage.

\subsection{Numerical Experiment}
We conduct a numerical study based on the IEEE 300-bus test system provided in MATPOWER \cite{zimmerman2011matpower}. The Laplacian $L$ is constructed from the bus-admittance matrix and the operating-point voltages in MATPOWER. At each bus, the generator dynamics are given by the standard swing equation with a first-order turbine model \cite{Jiang2021,Jiang2026FS}:
\begin{equation*}\label{eq: generator_dynamics}
g_i(s)=
\frac{\tau_i s+1}
{m_i\tau_i s^2+(m_i+d_i\tau_i)s+d_i+d_{t,i}},
\end{equation*}
where $m_i>0$ and $d_i>0$  are the aggregate generator inertia and damping, respectively, while $\tau_i>0$ and $d_{t,i}>0$ are the turbine time constant and inverse droop, respectively. We consider the inverters with iDroop \cite{Jiang2021,Jiang2026FS}:
\begin{equation*}\label{eq: inverter_iDroop}
c_i(s)=
-\frac{\nu_i s+\delta_i d_{b,i}}{s+\delta_i},
\end{equation*}
where $d_{b,i}>0$ is the inverter inverse droop, and $\nu_i,\delta_i>0$ are tunable parameters. Following the Nadir-elimination tuning in \cite{Jiang2021,Jiang2026FS}, we choose $\delta_i=\tau_i^{-1}$ and $\nu_i=d_{t,i}+d_{b,i}$, yielding the closed-loop bus dynamics
\begin{equation*}\label{eq: closed_loop_iDroop}
p_i(s)=\frac{g_i(s)}{1-g_i(s)c_i(s)}
=\frac{1}{m_i s+d_i+d_{t,i}+d_{b,i}}.
\end{equation*}

To introduce heterogeneity across bus dynamics, we independently sample $m_i$, $d_i$, $d_{t,i}$, and $d_{b,i}$ from uniform distributions $\mathcal{U}(8,12)$ s, $\mathcal{U}(3,5)$ p.u., $\mathcal{U}(4,6)$ p.u., and $\mathcal{U}(2,4)$ p.u., respectively. At each frequency $\omega$, we set $G(j\omega)=P(j\omega)$ and $H(j\omega)=\frac{L}{j\omega}$ so that the equation to be solved is
\[
    A_\omega x=b;  x=\begin{bmatrix}u_1\\u_2\end{bmatrix},  b=\begin{bmatrix}w_1\\0\end{bmatrix},
	A_\omega=
	\begin{bmatrix}
		I&H(j\omega)\\
		-G(j\omega)&I
	\end{bmatrix}.
\] 
The matrix $A_\omega \in \mathbb{C}^{600\times600}$ has $2016$ nonzero entries, corresponding to a density of $2016/600^2=0.56\%$.

\begin{figure}[htp]
	\centering
	\includegraphics[width=0.7\linewidth]{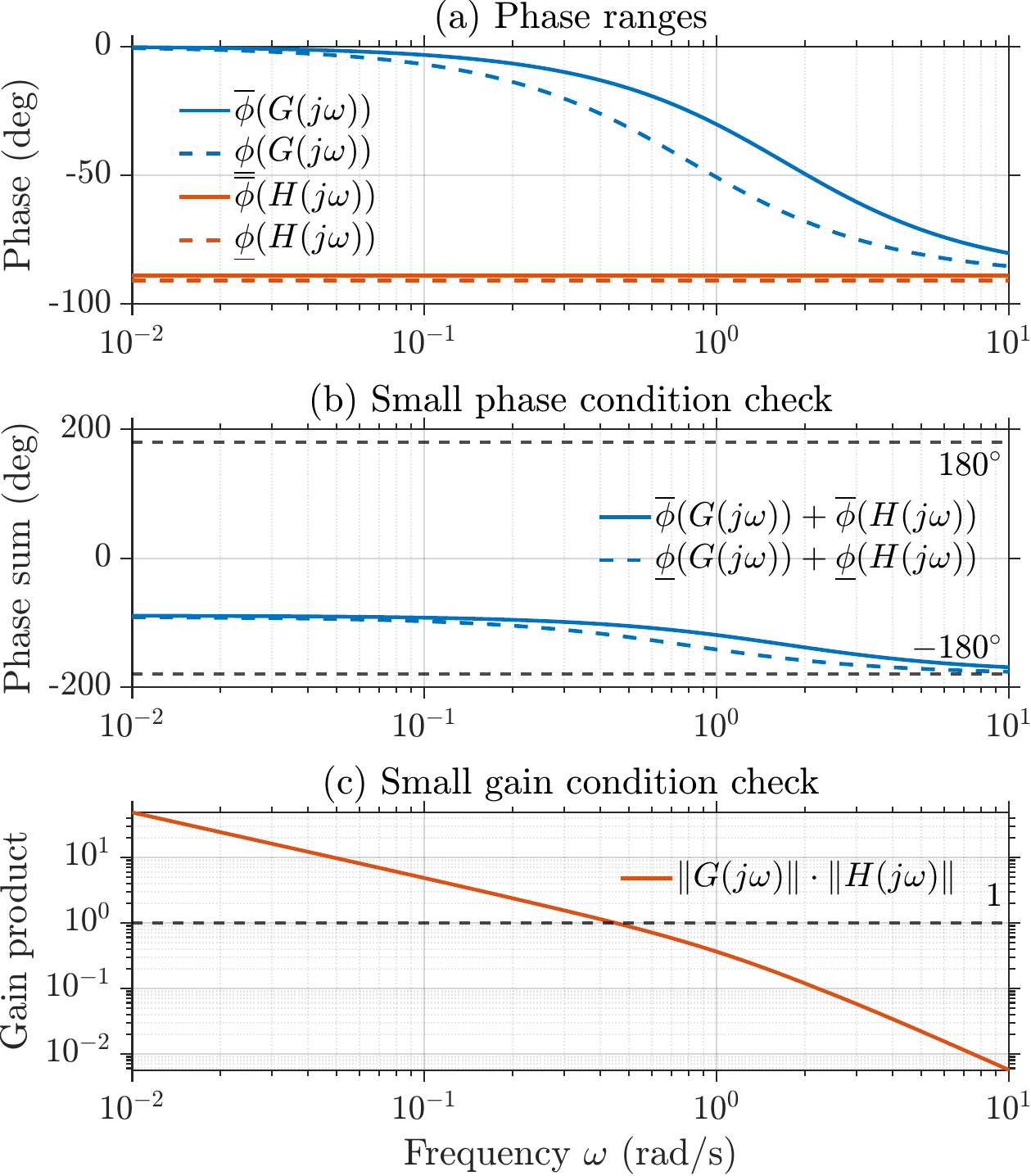}
	\caption{Applicability of small phase and
		small gain. (a) shows the phase ranges of
		$G(j\omega)$ and $H(j\omega)$; (b) reports the phase sums $\overline{\phi}(G(j\omega))
		+\overline{\phi}(H(j\omega))$ and $\underline{\phi}(G(j\omega))
		+\underline{\phi}(H(j\omega))$; (c) reports the gain product
		$\|G(j\omega)\| \cdot \|H(j\omega)\|$.}
	\label{fig:phase_gain_validation}
\end{figure}

We verify the small phase and small gain conditions over a dense frequency sweep $\omega\in[10^{-2},10^1]\ {\rm rad/s}$. As shown in Figure~\ref{fig:phase_gain_validation}, the small phase condition holds throughout the considered frequency range, whereas the small gain condition is not satisfied at low frequencies due to the large gain of $H(j\omega)$. Notably, the phases of $H(j\omega)$ remain constant over the entire positive frequency range.

\begin{figure}[htp]
	\centering
	\includegraphics[width=0.7\linewidth]{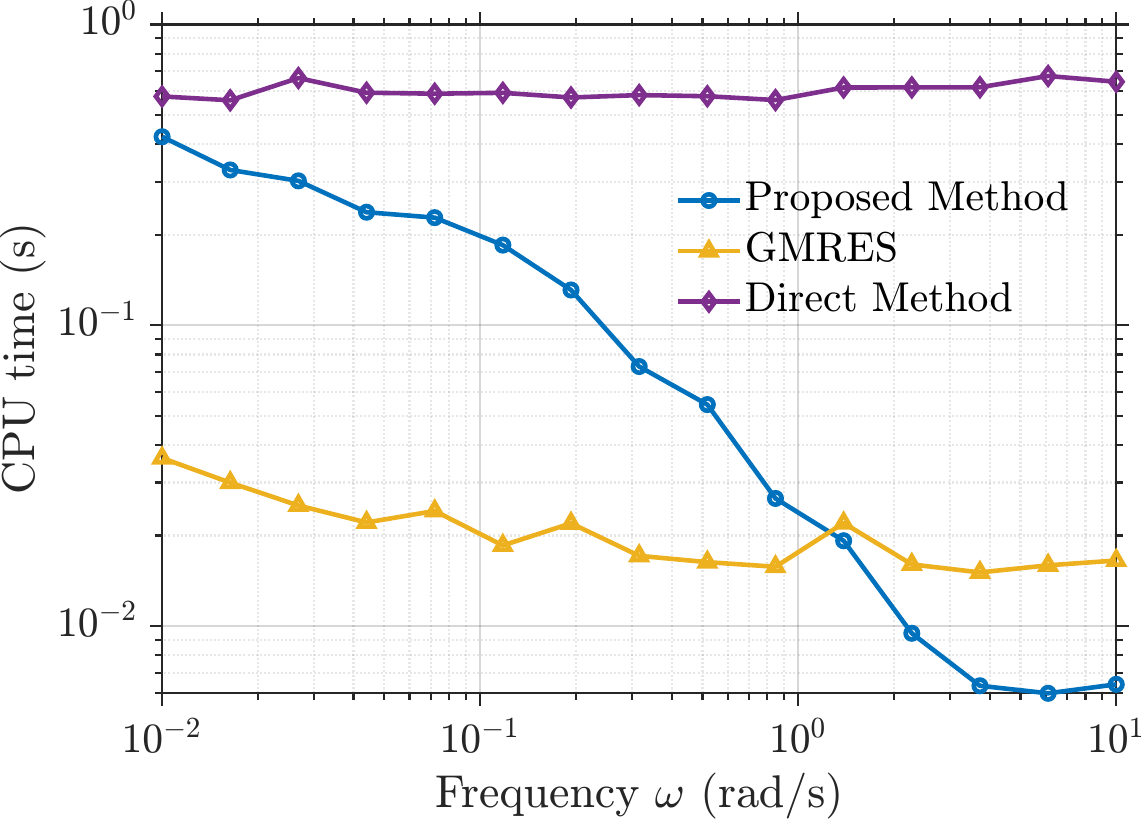}
	\caption{Solver CPU time of the proposed method, GMRES, and Gaussian elimination over the 15 test frequencies.}
	\label{fig:solver_comparison}
\end{figure}

For computational performance comparison, we consider GMRES as a representative Krylov subspace iterative method~\cite{greenbaum1997iterative} and Gaussian elimination with partial pivoting as a representative direct solver. The iterative methods are initialized at $x(0)=0$, and terminate when $\|b-A_\omega x(k)\|/\|b\|\leq10^{-8}$. The experiments are performed in MATLAB R2023a on a Windows workstation with an Intel Core i7-13650HX CPU and 32 GB of RAM. As shown in Figure~\ref{fig:solver_comparison}, the proposed method requires less solver CPU time than Gaussian elimination across the 15 test frequencies. Moreover, its solver time decreases with frequency and becomes lower than that of GMRES at higher frequencies, whereas the solver times of Gaussian elimination and GMRES remain relatively flat. 

\begin{figure}[htp]
	\centering
	\includegraphics[width=0.7\linewidth]{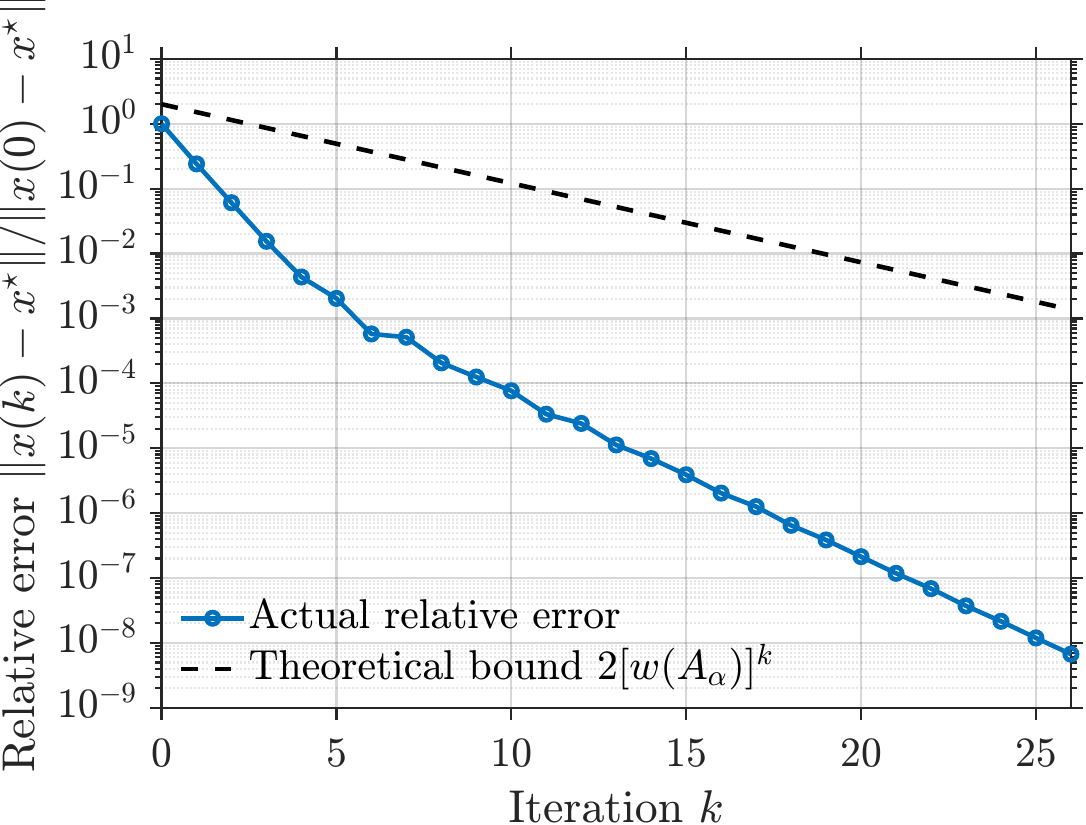}
	\caption{Actual relative error of the proposed method and the theoretical error bound $2[w(A_\alpha)]^k$ at $\omega=3.728\,{\rm rad/s}$.}
	\label{fig:residual_history}
\end{figure}

To examine the convergence behavior, we consider  $\omega=3.728\,{\rm rad/s}$ as a specific frequency. At this frequency,  we choose $\phi_0=-6.642^\circ$ from the admissible interval $[-12.310^\circ,-0.973^\circ]$ and set $\theta=0.8\theta_0=0.8$. The resulting matrix $T_\theta$ has $\gamma(T_\theta)=-0.253^\circ$ and $\delta(T_\theta)=75.470^\circ$. With step-size $\alpha=0.9\alpha_{\max}\approx 0.855$, the iteration matrix $A_\alpha=I-\alpha e^{-j\gamma(T_\theta)}T_\theta$ satisfies $w(A_\alpha)\approx0.756<1$. As shown in Fig.~\ref{fig:residual_history}, the observed relative error decays geometrically and remains below the theoretical bound established in Lemma~\ref{lemma: NR_contractive_convergence}.


\section{Conclusion}\label{section: conclusion}
In this paper, we developed an iterative method for solving feedback-type equations under small phase condition, by exploiting the phase information of related matrices and utilizing numerical-radius contraction as a convergence criterion.  A key technical step in developing the method is a preconditioning of equations via convex combination as in (\ref{Ttheta}), which transforms the feedback matrix into a strictly accretive form that enables convergent iterations even when classical gain-based methods fail. Interesting future works include refining parameter selection for fast convergence, exploring phase-based iterative methods in optimization problems, and extending the development to operators.

\appendix
\makeatletter
\def\@seccntformat#1{%
  \ifcsname appendixprefix@#1\endcsname
    \csname appendixprefix@#1\endcsname
  \fi
  \csname the#1\endcsname:\ 
}
\def\appendixprefix@section{Appendix~}
\makeatother

\section{Proof of Theorem~\ref{Thm: passive_iteration}}\label{appdix_proof: gain_phase_convex}
(i): A simple calculation shows that $\mathrm{Re}(T_{\theta})=\theta M$ with
\begin{equation*}
    M=\begin{bmatrix}
        I+\dfrac{1-\theta}{\theta}\mathrm{Re}(\widehat{G})& \dfrac{\widehat{H}-\widehat{G}^*}{2} \\
        \dfrac{\widehat{H}^*-\widehat{G}}{2} & I+\dfrac{1-\theta}{\theta}\mathrm{Re}(\widehat{H})
    \end{bmatrix}.
\end{equation*}
Clearly, for any $0<\theta<1$, $T_\theta$ is strictly accretive if and only if $M>0$. Let
\begin{equation*}
    \mathrm{Re}(\widehat{G})=U\cdot \mathrm{diag}\{\lambda_1(\mathrm{Re}(\widehat{G})),\cdots,\lambda_n(\mathrm{Re}(\widehat{G}))\}\cdot U^*,
\end{equation*}
\begin{equation*}
    \mathrm{Re}(\widehat{H})=V\cdot \mathrm{diag}\{\lambda_1(\mathrm{Re}(\widehat{H})),\cdots,\lambda_n(\mathrm{Re}(\widehat{H}))\}\cdot V^*,
\end{equation*}
be Schur decompositions of $\mathrm{Re}(\widehat{G})$ and $\mathrm{Re}(\widehat{H})$, where $U,V\in\mathbb{C}^{n\times n}$ are unitary matrices. Then,
\begin{align*}
    M>0 &\Leftrightarrow \begin{bmatrix}
        U & 0 \\
        0 & V
    \end{bmatrix}^* M \begin{bmatrix}
        U & 0 \\
        0 & V
    \end{bmatrix}>0 \\
    &\Leftrightarrow \begin{bmatrix}
        D_{\widehat{G}} & U^*\dfrac{\widehat{H}-\widehat{G}^*}{2}V \\
        V^*\dfrac{\widehat{H}^*-\widehat{G}}{2}U & D_{\widehat{H}}
    \end{bmatrix}>0,
\end{align*}
where $D_{\widehat{G}}=I+\dfrac{1-\theta}{\theta}\mathrm{diag}\{\lambda_1(\mathrm{Re}(\widehat{G})),\cdots,\lambda_n(\mathrm{Re}(\widehat{G}))\}$ and
$D_{\widehat{H}}=I+\dfrac{1-\theta}{\theta}\mathrm{diag}\{\lambda_1(\mathrm{Re}(\widehat{H})),\cdots,\lambda_n(\mathrm{Re}(\widehat{H}))\}$. Since $\widehat{G}, \widehat{H}$ are accretive, it follows that $D_{\widehat{G}}>0,D_{\widehat{H}}>0$. Then, an application of Schur complement lemma yields
\begin{align}
    M>0 &\Leftrightarrow D_{\widehat{H}}-V^*\dfrac{\widehat{H}^*-\widehat{G}}{2}UD_{\widehat{G}}^{-1}U^*\frac{\widehat{H}-\widehat{G}^*}{2}V>0 \nonumber \\
    &\Leftrightarrow \left\Vert D_{\widehat{G}}^{-\frac{1}{2}}U^*\frac{\widehat{H}-\widehat{G}^*}{2}VD_{\widehat{H}}^{-\frac{1}{2}} \right\Vert <1.\label{ine}
\end{align}
Hence, to show (i), it suffices to show inequality (\ref{ine}). To this end, from the sub-multiplicity and unitary invariance of norm $\Vert\cdot\Vert$, we have
\begin{align*}
&\left\Vert D_{\widehat{G}}^{-\frac{1}{2}}U^*\frac{\widehat{H}-\widehat{G}^*}{2}VD_{\widehat{H}}^{-\frac{1}{2}} \right\Vert \\
   & \hspace{10pt}\leq \left\Vert D_{\widehat{G}}^{-\frac{1}{2}} \right \Vert\cdot \left\Vert U^*\frac{\widehat{H}-\widehat{G}^*}{2}V\right\Vert\cdot \left\Vert D_{\widehat{H}}^{-\frac{1}{2}}\right\Vert \\
   &\hspace{10pt}=\frac{1}{\sqrt{1+\frac{1-\theta}{\theta}\lambda_{\min}(\mathrm{Re}(\widehat{G}))}} \cdot \left\Vert \frac{\widehat{H}-\widehat{G}^*}{2}\right\Vert \\
   &\hspace{10pt}\qquad \qquad \qquad \qquad \qquad\qquad\cdot \frac{1}{\sqrt{1+\frac{1-\theta}{\theta}\lambda_{\min}(\mathrm{Re}(\widehat{H}))}} \\
   &\hspace{10pt} \leq \frac{\left\Vert \frac{\widehat{H}-\widehat{G}^*}{2}\right\Vert}{\sqrt{1+\frac{1-\theta}{\theta}\left(\lambda_{\min}(\mathrm{Re}(\widehat{G}))+\lambda_{\min}(\mathrm{Re}(\widehat{H}))\right)}}<1
\end{align*}
for any $0<\theta<\theta_0$. This completes the proof of (i).

(ii): The convergence of iteration (\ref{eq: passive_iteration}) and the corresponding error bound follow from Lemma \ref{lemma: sectorial iteration} as $T_\theta$ is strictly accretive. We only need to show the solution of (\ref{eq: core_feedback_equation}) and the solution of 
\begin{equation}\label{eq: T_theta equation}
    T_\theta \begin{bmatrix}
        \widehat{u}_1 \\
        \widehat{u}_2
    \end{bmatrix}=\begin{bmatrix}
    \theta\widehat{w}_1-(1-\theta)\widehat{w}_2\\
    (1-\theta)\widehat{w}_1+\theta\widehat{w}_2
\end{bmatrix}
\end{equation}
are the same. This is straightforward since (\ref{eq: T_theta equation}) can be obtained from multiplying both sides of (\ref{eq: core_feedback_equation}) by the nonsingular matrix
$
\begin{bmatrix}
    \theta I & -(1-\theta)I \\
    (1-\theta)I & \theta I
\end{bmatrix}.
$
This completes the proof of (ii).

\section{Parameter Selection} \label{appdix: parameter_selection}
\textit{Phase rotation $\phi_0$.} Recall that the admissible interval for $\phi_0$ is characterized by \eqref{eq: phi0_select_condition}. If one of $G$ and $H$ is semi-sectorial with field angle equal to $\pi$, say $H$, then $\phi_0=\underline{\phi}(H)+\frac{\pi}{2}=\overline{\phi}(H)-\frac{\pi}{2}$ is uniquely determined. 

When neither $G$ nor $H$ is semi-sectorial with field angle equal to $\pi$, denote by $\mathcal{I}_{\phi_0}$ the interval of admissible phase rotation. Recall that the upper bound $\theta_0$ of later convex-combination parameter $\theta$ is given by \eqref{theta0}. Note that $\widehat{H}-\widehat{G}^*
=e^{-j\phi_0}(H-G^*),$ and hence $\|\widehat{H}-\widehat{G}^*\|$ is independent of $\phi_0$. Hence, selecting $\phi_0$ to enlarge the admissible range of $\theta$ and thereby admit potentially faster convergence is desirable. Maximizing $\theta_0$ (when $\|(\widehat{H}-\widehat{G}^*)/2\|>1$) amounts to 
\begin{align*}
    \text{maximize } &\lambda_{\min}(\mathrm{Re}(\widehat{G})) + \lambda_{\min}(\mathrm{Re}(\widehat{H}))\\
    \text{subject to }&\phi_0\in \mathcal{I}_{\phi_0}.
\end{align*}
This is a convex optimization problem. To see this, let $x$ be a unit vector, define
\[
g_x(\phi_0)
=x^*\mathrm{Re}(\widehat{G})x
=\mathrm{Re}(e^{j\phi_0}x^*Gx).
\]
Differentiating twice gives
$g_x''(\phi_0)=-g_x(\phi_0)$.
Since
$\mathrm{Re}(\widehat{G})\geq 0$, we have
$g_x(\phi_0)\ge0$, so $g_x''(\phi_0)\le0$.
Then every $g_x$ is concave on $\mathcal{I}_{\phi_0}$. Since $\lambda_{\min} (\mathrm{Re}(\widehat{G}))
=\min_{\|x\|=1}g_x(\phi_0)$ and pointwise infimum of concave functions is concave, we know $\lambda_{\min} (\mathrm{Re}(\widehat{G}))$ is concave on $\mathcal{I}_{\phi_0}$. Similarly for $\lambda_{\min} (\mathrm{Re}(\widehat{H}))$. This shows that choosing phase rotation $\phi_0$ to maximize the upper bound of convex combination parameter $\theta$ is a convex problem.

From a geometric perspective, maximizing the sum of $\lambda_{\min}(\mathrm{Re}(\widehat{G}))$ and $\lambda_{\min}(\mathrm{Re}(\widehat{H}))$ motivates a heuristic choice of $\phi_0$ near the center of the admissible phase rotation interval $\mathcal{I}_{\phi_0}$.

\textit{Convex-combination parameter $\theta$ and step size $\alpha$}. Our preliminary analysis suggests that the choice of $\theta$ and $\alpha$ is closely related to the geometry of $W(e^{-j\gamma(T_\theta)}T_\theta)$. Denote by $B_{\theta}=e^{-j\gamma(T_\theta)}T_\theta$ and
\[
\nu_\theta=\lambda_{\min} (\mathrm{Re}(B_\theta)),\quad \mu_\theta=\lambda_{\max} (\mathrm{Re}(B_\theta)).
\]
Then $W(B_\theta)$ lies inside a trapezoid with vertices
\[
\nu_\theta \pm j\nu_\theta \tan \frac{\delta(T_\theta)}{2}, \mu_\theta \pm j\mu_\theta \tan \frac{\delta(T_\theta)}{2}.
\]
Hence
\begin{align*}
w(I-\alpha B_\theta)^2&=\max_{a+jb\in W(B_\theta)}\left[(1-\alpha a)^2+\alpha^2b^2\right]\\
&\leq \max_{\nu_\theta\leq a \leq \mu_\theta}\left[1-2\alpha a+\alpha^2a^2\sec^2\frac{\delta(T_\theta)}{2}\right],
\end{align*}
where the second maximum is attained at $a=\nu_\theta$ or $a=\mu_\theta$. This gives an upper bound of $w(I-\alpha B_\theta)$, denoted by 
\[
q(\alpha)=\sqrt{\max_{\nu_\theta\leq a \leq \mu_\theta}\left[1-2\alpha a+\alpha^2a^2\sec^2\frac{\delta(T_\theta)}{2}\right]}.
\]
Minimizing $q(\alpha)$ over step size $\alpha$ gives
\begin{align}
    \alpha^\star&=\frac{2\cos^2 \frac{\delta(T_\theta)}{2}}{\nu_\theta+\mu_\theta}, \text{ and}\label{alpha}\\
    q(\alpha^*)&=\sqrt{1-\frac{4\nu_\theta \mu_\theta}{(\nu_\theta+\mu_\theta)^2}\cos^2\frac{\delta(T_\theta)}{2}}.\label{q}
\end{align}
The expression (\ref{q}) suggests that it is favorable to select $\theta$ so as to make both $\frac{\mu_\theta}{\nu_\theta}$ (condition number of $\mathrm{Re}(B_\theta$)) and $\delta(T_\theta)$ (field angle of $T_\theta$) small. There may be a trade-off between these two in general. 

%An illustrative example is given in Figure~\ref{fig: theta_selection}. Once $\theta$ is selected, the expression (\ref{alpha}) gives a candidate choice of step size.

% \begin{figure}[htp]
%     \centering
%     \begin{subfigure}{0.35\linewidth}
%         \centering
%         \includegraphics[width=\linewidth]{theta0_selection_NR.pdf}
%         \caption{Numerical ranges of $B_\theta$.}
%     \end{subfigure}
%     \hfil
%     \begin{subfigure}{0.45\linewidth}
%         \centering
%         \includegraphics[width=\linewidth]{theta0_selection_tradeoff.pdf}
%         \caption{Variation of $\mu_\theta/\nu_\theta$ and
%         $\delta(T_\theta)$.}
%     \end{subfigure}
% \caption{
% Effect of the convex-combination parameter $\theta$ on
% the numerical range of
% $B_\theta=e^{-j\gamma(T_\theta)}T_\theta$.
% The condition number $\mu_\theta/\nu_\theta$ and the field angle $\delta(T_\theta)$ trade off as $\theta$ varies. Here, $G$ and $H$ are the same as those in Fig.~\ref{fig: case02_phi0_selection}, with $\phi_0=-120^\circ$.
% }
%     \label{fig: theta_selection}
% \end{figure}

% \begin{ack}                               % Place acknowledgements
% This work was supported in part by the National Natural Science Foundation of China under grants 62573006 and 72131001, by the Natural Science Foundation of Guangdong Province under grants 2024A1515011630, and by the Research Grants Council of Hong Kong under the General Research Fund No. 16206324.  % here.
% \end{ack}

\bibliographystyle{plain}        % Include this if you use bibtex 
\bibliography{autosam}           % and a bib file to produce the 
                                 % bibliography (preferred). The
                                 % correct style is generated by
                                 % Elsevier at the time of printing.

%\begin{thebibliography}{99}     % Otherwise use the  
                                 % thebibliography environment.
                                 % Insert the full references here.
                                 % See a recent issue of Automatica 
                                 % for the style.
%  \bibitem[Heritage, 1992]{Heritage:92}
%     (1992) {\it The American Heritage. 
%     Dictionary of the American Language.}
%     Houghton Mifflin Company.
%  \bibitem[Able, 1956]{Abl:56}
%     B.~C.~Able (1956). Nucleic acid content of macroscope. 
%     {\it Nature 2}, 7--9. 
%  \bibitem[Able {\em et al.}, 1954]{AbTaRu:54}   
%     B.~C. Able, R.~A. Tagg, and M.~Rush (1954).
%     Enzyme-catalyzed cellular transanimations.
%     In A.~F.~Round, editor, 
%     {\it Advances in Enzymology Vol. 2} (125--247). 
%     New York, Academic Press.
%  \bibitem[R.~Keohane, 1958]{Keo:58}
%     R.~Keohane (1958).
%     {\it Power and Interdependence: 
%     World Politics in Transition.}
%     Boston, Little, Brown \& Co.
%  \bibitem[Powers, 1985]{Pow:85}
%     T.~Powers (1985).
%     Is there a way out?
%     {\it Harpers, June 1985}, 35--47.

%\end{thebibliography}

        % Sections and subsections are supported  
                                        % in the appendices.
\end{document}